\documentclass[cameraready]{Interspeech}
\newcommand{\ours}{\mbox{\textsc{AWM}}\xspace}

\usepackage{subfigure}
\usepackage{textcomp}
\usepackage{xcolor}
\usepackage{graphicx}
\usepackage{multirow}
\usepackage{textcomp}
\usepackage{tikz}
\usepackage{xcolor}
\usepackage{hyperref}
\usepackage{xspace}
\usepackage{listings}
\usepackage{bbding}
\usepackage{wasysym}
\usepackage{enumitem,kantlipsum}
\usepackage{textcomp}
\usepackage{xcolor}
\usepackage{colortbl}
\usepackage{soul}
\usepackage{interval}
\usepackage{pgfplots}
\pgfplotsset{compat=1.18}
\usepackage{pifont}
\usepackage{url}
\usepackage{phonetic}
\usepackage{interval}
\usepackage{pgfplots}
\usepackage{tipa}
\usepackage{multicol}
\usepackage{multirow}
\usepackage{xspace}
\usepackage{xfrac}
\usepackage{pifont}
\usepackage{booktabs}
\usepackage{algorithm, algorithmicx}
\usepackage[noend]{algpseudocode}
\usepackage{amsmath}

\title{Learning to Evade: Adaptive Attacks on Audio Watermarking}

\author[affiliation={1}]{Weikang}{Ding$^{*}$}
\author[affiliation={2}]{Hanqing}{Guo}
\author[affiliation={1}]{Rui}{Duan}
\author[affiliation={3}]{Guangjing}{Wang}
\author[affiliation={4}]{Yuanda}{Wang}
\author[affiliation={5}]{Mingzhe}{Chen}
\author[affiliation={4}]{Qiben}{Yan}

\address{
    $^1$ University of Missouri-Kansas City, USA \\
    $^2$ University of Hawaii at Manoa, USA \\
    $^3$ University of South Florida, USA \\
    $^4$ Michigan State University, USA \\
    $^5$ University of Miami, USA
}

\email{wdhhd@umkc.edu, qyan@msu.edu}


\usepackage{comment}

\begin{document}

\maketitle

\begingroup
\renewcommand{\thefootnote}{*}
\footnotetext{Research was performed at Michigan State University.}
\endgroup

\begin{abstract}
Advances in generative audio have intensified copyright concerns, making audio watermarking increasingly important for asserting ownership.
However, existing audio watermarking methods are vulnerable to adversarial attacks. 
We find that watermark decoder message probabilities follow normal distributions, a property exploited by defenses to detect manipulations. 
This paper introduces an adaptive audio watermark attack method (AWM) designed to bypass existing defense strategies. \ours uses a two-stage optimization: the first stage ensures attack success, while the second improves audio quality. To evade detection, it estimates normal distribution parameters from limited samples of the target audio, and then adaptively steers decoded probabilities back into the estimated range. Evaluated on two watermarking methods across three voice datasets, \ours achieves high success while bypassing state-of-the-art detectors: detection rates are below 10\% for replacement and creation, and 0\% for removal.
\end{abstract}

\section{Introduction}\label{sec:introduction}

In recent years, the rapid growth of social networking platforms has encouraged many users to publicly share their audio content, including original works such as audiobooks and self-produced music. These audio contents might bring them income. However, many unauthorized users copy creative works, modify them, and re-upload them to mainstream platforms for profit, which significantly discourages original audio creators. Besides, voice cloning attacks can illegally synthesize the target’s voice for malicious purposes, potentially resulting in severe consequences such as financial losses and reputation damage~\cite{macfeereport}.

To address these issues, deep-learning-based watermarking has been proposed~\cite{ren2024copyright}. It embeds a noise-tolerant signal into the target audio, which remains imperceptible to human hearing while being detectable by specialized AI models. Accordingly, most approaches adopt an encoder–decoder architecture to embed and recover the signal. To improve robustness, training pipelines incorporate training data from distortion attack models, e.g., re-recording~\cite{liu2023dear}, voice cloning~\cite{timbrewatermarking-ndss2024, sanroman2024proactive}, and lossy codec compression~\cite{zhou2025wmcodec}, such that the watermark could survive common real-world transformations.

\begin{figure}
    \centering
    \includegraphics[width=1\linewidth]{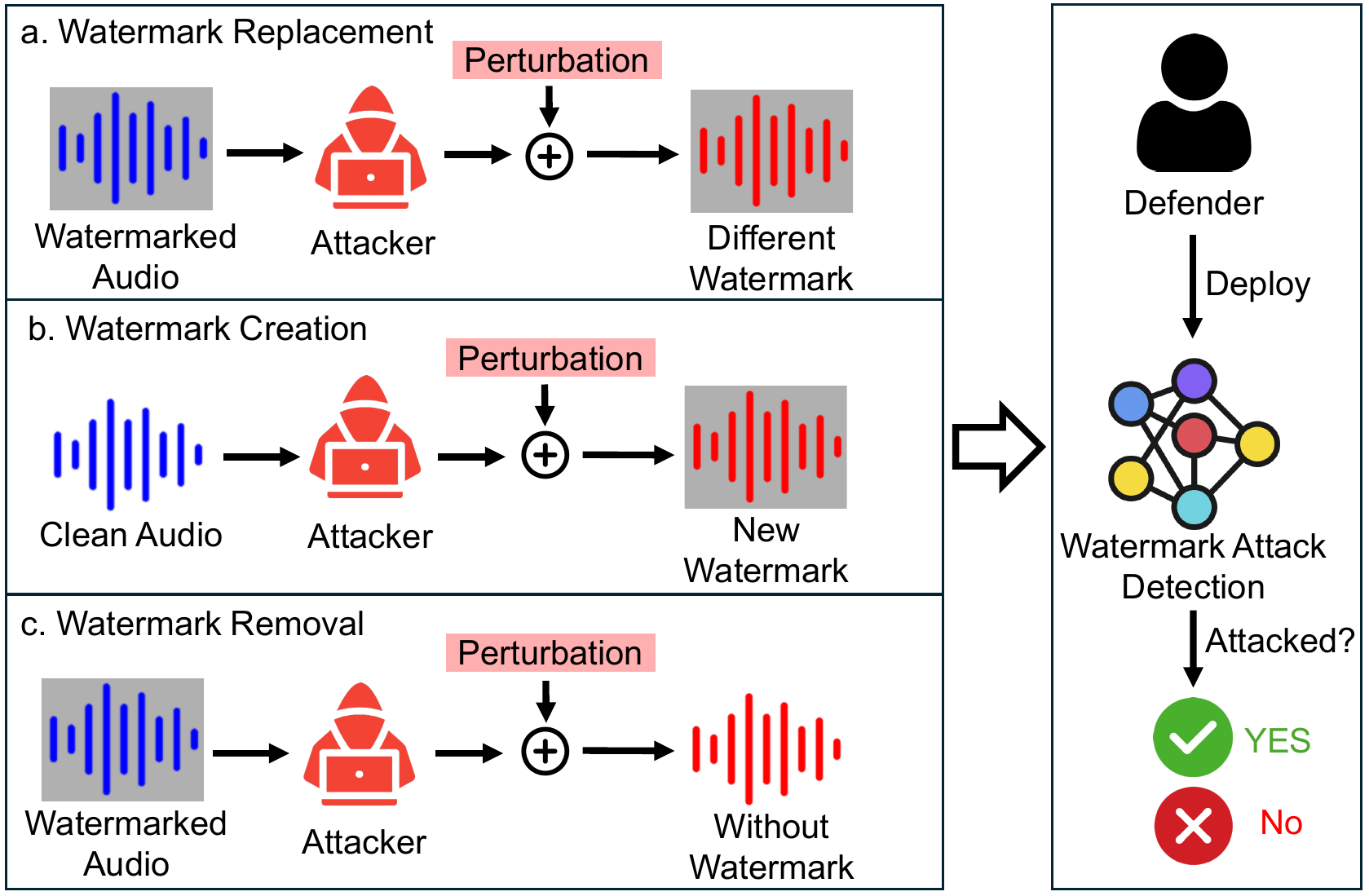}
    \vspace{-15pt}
    \caption{Overview of the watermark attack (left) and the watermark attack detection process used to detect whether the audio has been tampered with (right).}
    \label{fig:cover}
    \vspace{-10pt}
\end{figure}

However, the robustness of deep-learning-based audio watermarking methods against adversarial attacks remains a critical concern~\cite{wen2025sok}. Recent studies~\cite{liu2024audiomarkbench, yang2024can} show that attackers can remove or forge watermarks by embedding carefully crafted adversarial perturbations. These attacks operate by adding and optimizing a perturbation signal so that the watermark decoder is misled into producing incorrect outputs. Yet, existing attack strategies face the following challenges:

\emph{\textbf{C1:} 
How to design an attack method to balance audio quality and attack effectiveness?} The encoder ensures the watermark is imperceptible, while the decoder recovers the watermark under various distortions. In practice, the decoder is more publicly accessible and therefore becomes the primary source of model knowledge in most attack scenarios. However, deceiving the decoder may introduce certain noise artifacts. An overly aggressive attack can significantly degrade audio quality, while prioritizing perceptual quality may compromise attack success. As a result, it is essential to strike a balance that ensures sufficient attack effectiveness while preserving audio quality.

\emph{\textbf{C2:} How to design an attack to bypass the detection strategy?} For a given audio input, the decoder outputs both a binary message and the associated bit-wise probabilities. These probabilities tend to follow a normal distribution, and defenders can leverage this property to detect whether an audio sample has been tampered with. Consequently, after successfully altering the binary watermark message bits, attackers must further optimize the audio to ensure that the decoded message probabilities fall within the range classified by the defender as non-outliers.

\emph{\textbf{C3:} How to select the suitable audio samples to estimate the decoded message probability distribution?} To estimate the distribution parameters required for a successful and stealthy attack, they must instead rely on querying and analyzing the decoder’s outputs. Because different input samples induce different probability patterns, the parameters estimated by the attacker can diverge significantly from the defender’s true parameters, and this mismatch can ultimately lead to attack failure. Therefore, it necessitates designing an effective estimation strategy to improve the attack success rate.





In this paper, we propose \ours, an Adaptive audio WaterMark attack method, which is capable of bypassing the defender's detection strategy. Figure~\ref{fig:cover} illustrates the application scenarios. The attacker obtains the target audio and adds an adversarial perturbation to generate the perturbed audio. The defender then receives the perturbed audio and uses the watermark decoder to extract the message probabilities. A predefined set of distribution parameters is employed to detect outliers. If any outliers are identified, the audio is classified as ``attacked"; otherwise, it is considered ``clean". We propose three schemes to address the challenges above.




To address \textbf{C1},  we design a two-step attack framework. The first step focuses on maximizing attack effectiveness while evading the defender’s detection strategy. The second step aims to improve audio quality. Inspired by~\cite{lukas2022sok}, we introduce a threshold-based post-processing step that refines the perturbation to enhance audio quality, while constraining the decoded message probabilities to remain within the expected normal range.

To address \textbf{C2}, we develop an adaptive optimization strategy that explicitly steers the decoded message probabilities toward the estimated normal distribution range. The key idea is to prioritize optimization on probabilities that fall outside this range: if a perturbed bit probability already lies within the estimated normal region, its optimization weight is reduced. This yields a dynamic optimization process that concentrates on the most ``at-risk" binary message bits.

To address \textbf{C3}, we leverage the observation that data with similar feature distributions tend to produce similar decoded message probability distributions~\cite{pan2023asset, pan2024finding}. Accordingly, we estimate the required probability distribution parameters by collecting a limited set of audio samples whose features closely resemble those of the target audio.


In this paper, we make the following contributions:
\begin{itemize}[topsep=0pt]
\item We empirically demonstrate that the decoded message probabilities output by the watermark decoder tend to follow normal distributions. This statistical property can be leveraged by defenders to design detection strategies based on outlier detection. 

\item We propose \ours, an adaptive audio watermark attack meth\-od that supports three attack types. Through a two-step optimization framework, \ours can bypass distribution-based detection strategies and preserve audio quality. The adaptive optimization strategy focuses updates on the message bits with the most uncertain probabilities, further improving attack effectiveness.



\item We evaluate \ours on three speech datasets and two state-of-the-art watermarking models. Compared to baseline methods, \ours achieves superior performance in both Attack Success Rate (ASR) and Detection Success Rate (DSR). Moreover, even after applying five no-box perturbations, AWM consistently maintains a high ASR, with most scores approaching or reaching 100\%.
\textbf{The source code and demos are available on} \url{https://adaptiveaudiowmattack.github.io/}


\end{itemize}

\section{Related Work}\label{sec:relatedwork} 

\noindent\textbf{Deep Learning-Based Audio Watermarking.} Unlike traditional schemes~\cite{hua2016twenty}, which rely on predefined transformations, deep-learning-based schemes can learn complex feature representations and optimize watermarking dynamically~\cite{timbrewatermarking-ndss2024, sanroman2024proactive, chen2023wavmark, juvela2025audio, singh2024silentcipher}. These schemes follow the architecture of the Encoder-Distortion-Decoder. The encoder embeds the watermark into the audio and generates the watermarked audio, the decoder receives the watermarked audio and extracts the corresponding watermark, and the distortion simulates a variety of potential attack scenarios. This paper implements the defense mechanism and attack strategy by utilizing this scheme.

\noindent\textbf{Audio Watermark Attack.} Audio watermark attacks based on adversarial perturbations can be categorized as no-box, black-box, or white-box, depending on the attacker's knowledge of the watermarking model.
In no-box perturbations, the attacker relies on generic audio processing techniques (e.g., compression~\cite{defossez2022high}, voice conversion~\cite{chou2019one,lin2021fragmentvc}, and text-to-speech~\cite{shen2018natural, ren2020fastspeech}). In black-box perturbations, the attacker can query the watermarking detector but does not have access to its internal architecture or parameters~\cite{chen2020hopskipjumpattack, andriushchenko2020square} .
In white-box perturbations, the attacker has full access to the detector, including its architecture and parameters, and introduces a perturbation to perform the gradient-based attack~\cite{carlini2017towards, madry2017towards}. While some existing studies~\cite{yang2024can, jiang2023evading, liu2024audiomarkbench} are either easily detected or generate the perturbed audio with poor quality. In this paper, we focus on balancing the attack effectiveness and audio quality.

\noindent\textbf{Audio Watermark Attack Detection.} Adversarial examples and watermarks both achieve the desired goal by adding imperceptible noise. Different from adversarial examples, adding watermarks to objects has little impact on the performance of the model inference (such as audio classification models~\cite{elizalde2023clap} and speech recognition models~\cite{radford2023robust}). Therefore, some outlier detection methods based on the time series~\cite{blazquez2021review, wu2022timesnet} are ineffective for identifying whether an audio sample has been tampered with. Recent studies have explored detection methods for generated images~\cite{wang2023dire, ha2024organic}, watermark images~\cite{pan2024finding, jiang2023evading}, and audio deepfake~\cite{afchar2024detecting, zang2024singfake}. However, to the best of our knowledge, these approaches cannot be directly applicable to detecting audio watermark attacks.

\section{Background}\label{sec:bk}



\subsection{Preliminary}
Adding perturbations to the audio is a common strategy for attacks against watermarking systems. The core idea is to either destroy the original watermark or forge a new one by introducing perturbations that deceive the watermark decoder.

\noindent\textbf{Audio Watermark Decoder.} The audio watermark decoder $Dec(\cdot)$ takes the encoded audio as input and outputs the extracted message. The extracted message can be represented in two forms: a binary message and message probabilities. The binary message is a direct representation of the decoded output as a sequence of 0s and 1s, denoted as $m \in \{0,1\}^N$. The message probabilities provide the decoder’s confidence values for each bit, indicating the likelihood of the bit being 1 or 0. A predefined threshold $\theta$ is used to convert probabilities into binary message values: if a probability exceeds the threshold, the binary message bit is decoded as 1; otherwise, it is decoded as 0. The form of message probabilities can be written as:
\begin{equation}
\begin{aligned}
    p=Dec(s),
    \label{eq:decoder}
\end{aligned}
\end{equation}
where $s$ is clean or watermarked audio, $Dec(\cdot)$ is the watermark decoder.

Let $s_{c}$ denote clean audio and $s_w$ denote watermarked audio. The watermark decoder $Dec(\cdot)$ extracts the clean message probabilities $p_{c}$ from $s_{c}$ and the watermarked message probabilities $p_w$ from $s_w$, respectively. The attacker specifies a target message probability, denoted as $p_{t}$. A perturbation $\delta$ is introduced to perform the attack on the audio.

\noindent\textbf{Watermark Replacement.}
In watermark replacement, a perturbation $\delta$ is added to the watermarked audio $s_w$, deceiving the decoder into misclassifying the embedded watermark as a different one. Formally, the goal of watermark replacement is:
\begin{equation}
    \delta_{replacement}=\arg\min\limits_{\delta}\lVert Dec(s_w+\delta)-p_{t}\rVert.
    \label{eq:watermark_replacement}
\end{equation}

\noindent\textbf{Watermark Creation.}
Watermark creation involves adding a perturbation $\delta$ to a clean audio $s_{c}$ to generate a perturbed watermarked audio, which deceives the watermark decoder into recognizing it as containing a valid watermark. Formally, the goal of watermark creation is:
\begin{equation}
    \delta_{creation}=\arg\min\limits_{\delta}\lVert Dec(s_{c}+\delta)-p_{t})\rVert.
    \label{eq:watermark_creation}
\end{equation}
\noindent\textbf{Watermark Removal.} Watermark removal is an untargeted attack aimed at removing the embedded watermark. We introduce a watermark detector, $Detector(\cdot)$, which determines whether the audio contains a watermark. A perturbation $\delta$ is added to the watermarked audio $s_w$ to mislead the detector into classifying the perturbed audio as unwatermarked (i.e., clean). Formally, the goal of watermark removal can be expressed as:
\begin{equation}
    \delta_{removal}=\arg\min\limits_{\delta}\lVert\delta\rVert\ \text{s.t.}\ Detector(s_w+\delta)=\text{None}.
    \label{eq:watermark_removal}
\end{equation}


\subsection{Message Probability Distribution} \label{subsec:message_probaility_distribution}
\begin{figure}[t]
    \centering
    
    \subfigure[Normal distribution with clean audio (Timbre).]{
        \label{fig:timbre_normal_distribution_a}
        \includegraphics[width=0.22\textwidth]{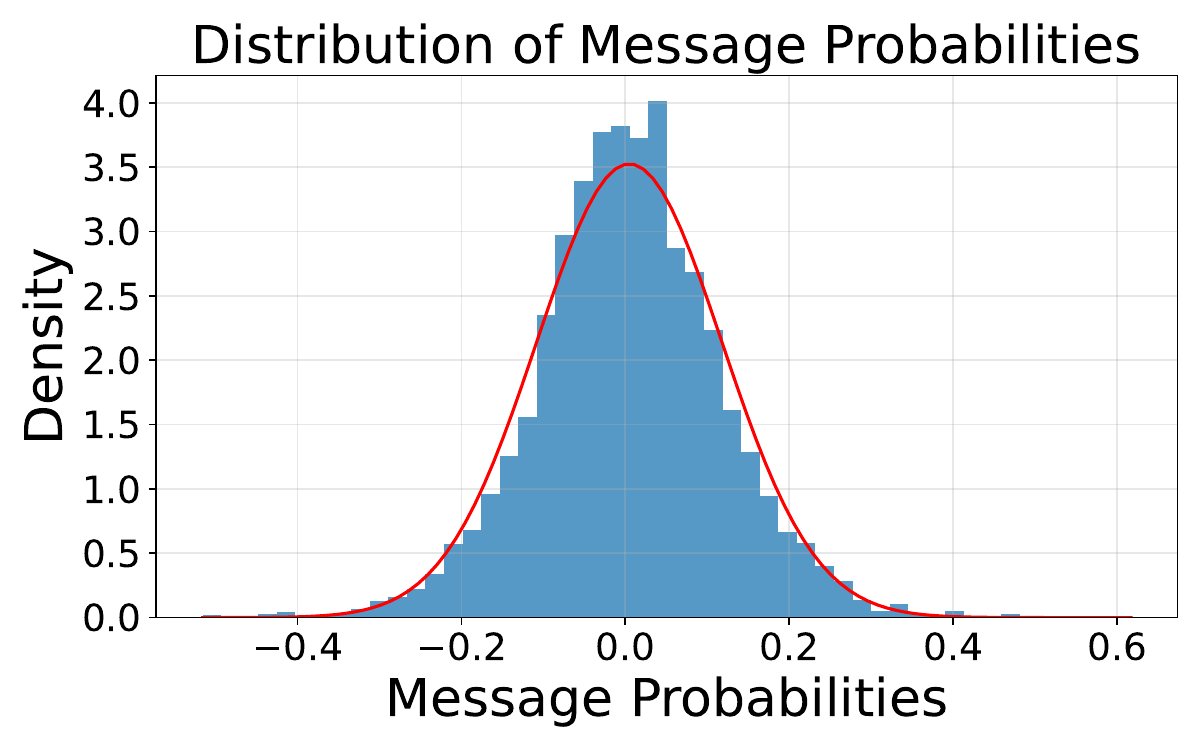}
    }
    \hspace{0.002\textwidth}
    \subfigure[Normal distribution with watermarked audio (Timbre).]{
        \label{fig:timbre_normal_distribution_b}
        \includegraphics[width=0.22\textwidth]{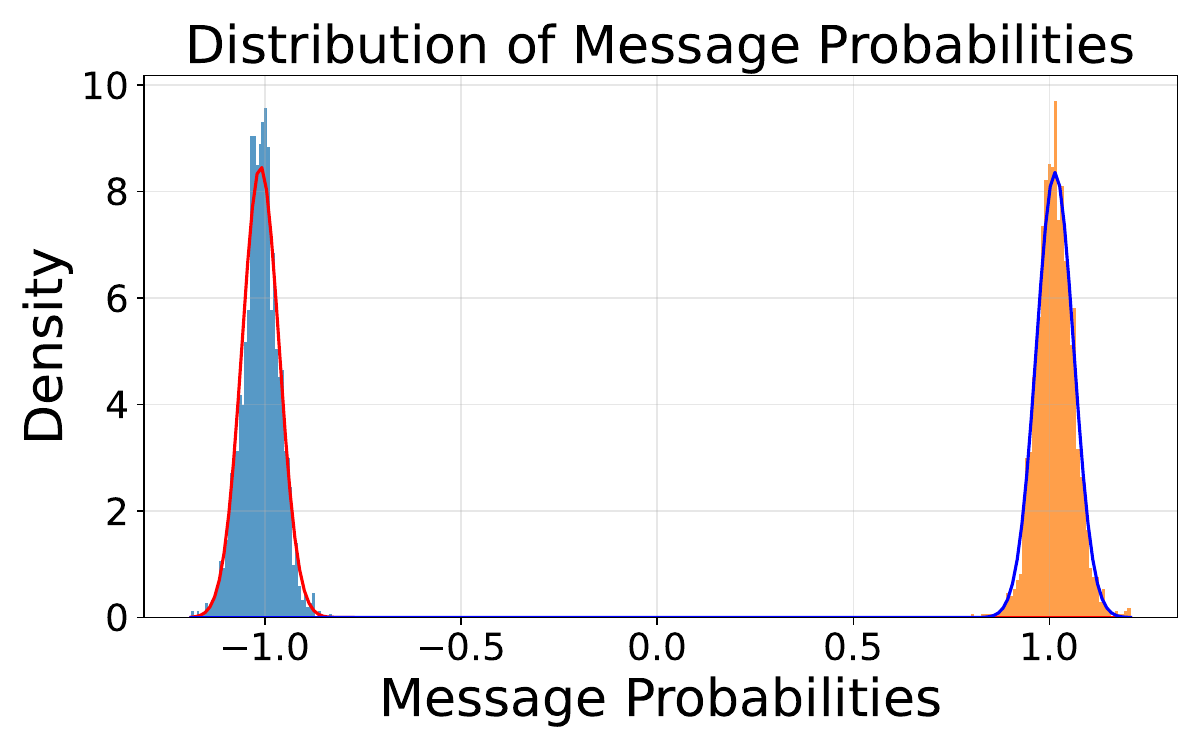}
    }
    \subfigure[Normal distribution with clean audio (AudioSeal).]{
        \label{fig:audioseal_normal_distribution_a}
        \includegraphics[width=0.22\textwidth]{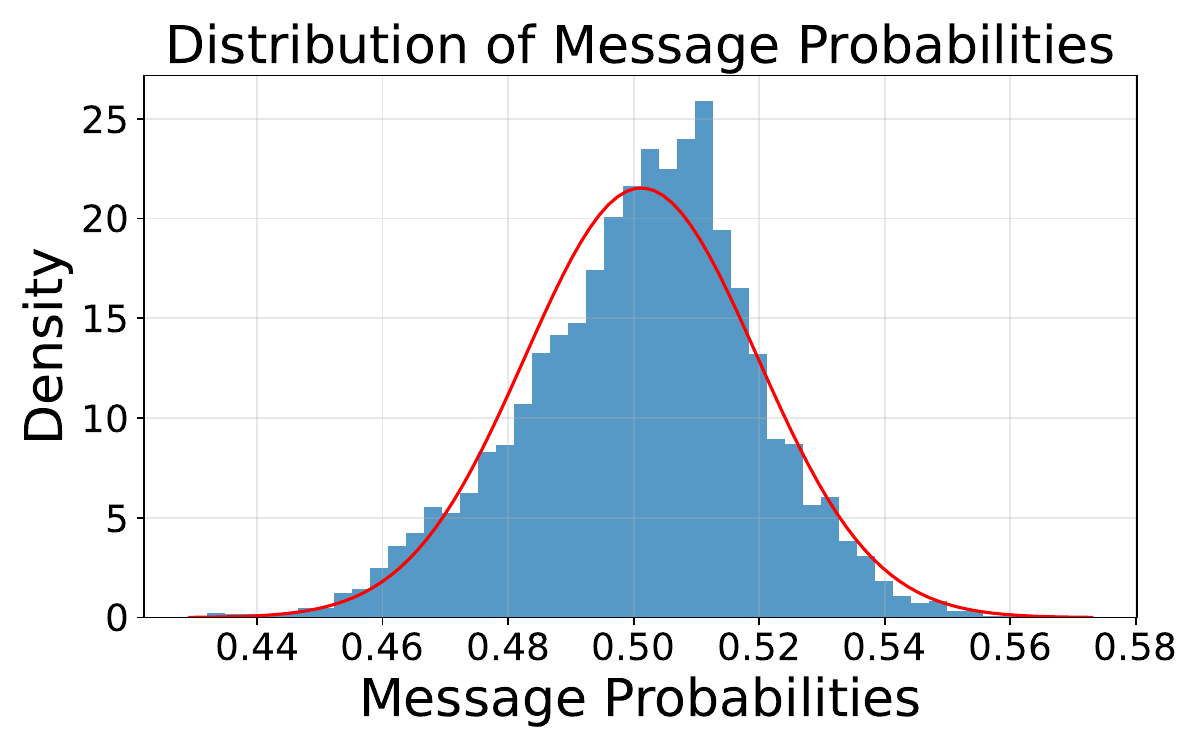}
    }
    \hspace{0.002\textwidth}
    \subfigure[Normal distribution with watermarked audio (AudioSeal).]{
        \label{fig:audioseal_normal_distribution_b}
        \includegraphics[width=0.22\textwidth]{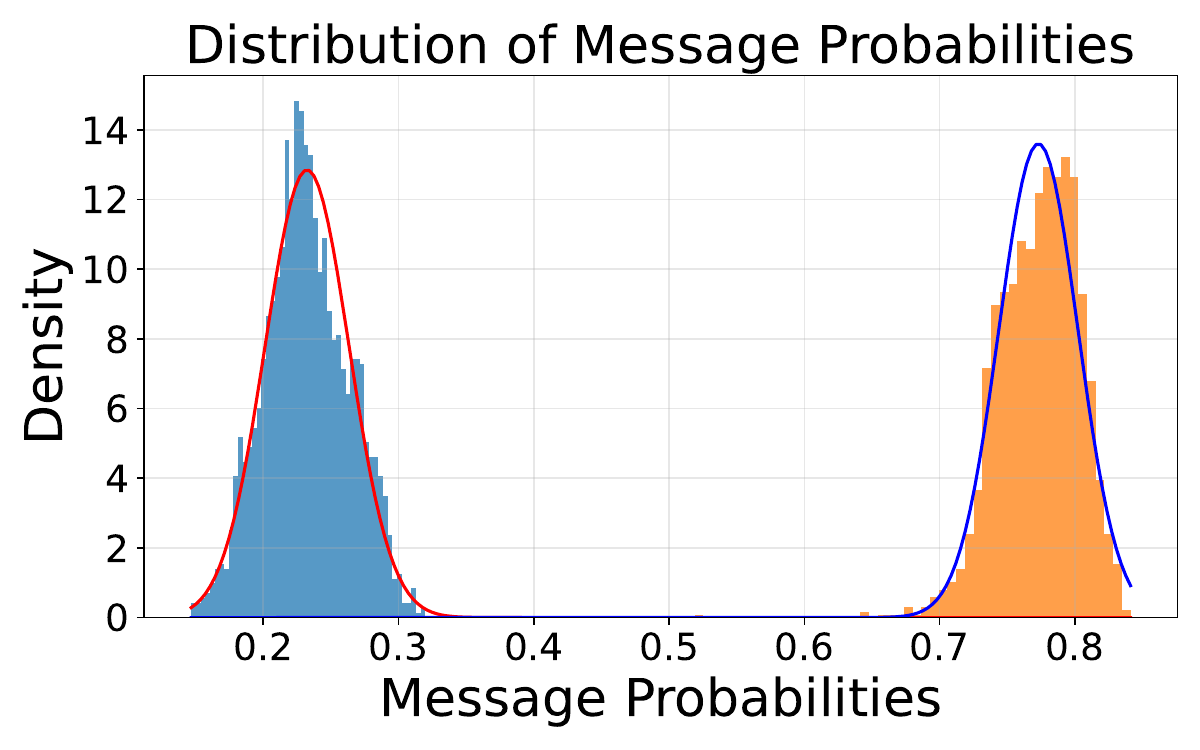}
    }
    \vspace{-3mm}
    \caption{Distribution of benign message probabilities.}
    \label{fig:normal_distribution_findings}
    \vspace{-10pt}
\end{figure}

\noindent\textbf{Benign Data Distribution.}
The watermark decoder outputs the message probabilities for each binary message bit. A bit is decoded as 1 if its probability exceeds a predefined threshold; otherwise, it is decoded as 0. For both clean and watermarked audio, we observe two distinct distributions, each following a normal distribution pattern. Clean audio exhibits a unimodal normal distribution, with the peak density centered near the predefined threshold. Specifically, in the Timbre, message probabilities below 0 are decoded as 0, while message probabilities above 0 are decoded as 1. As shown in Figure \ref{fig:timbre_normal_distribution_a}, where the threshold is set to 0, the mean $\mu$ of the distribution is also close to 0. In contrast, the watermarked audio follows a bimodal, approximately normal distribution (Figure \ref{fig:timbre_normal_distribution_b}), with two peaks corresponding to the decoded binary message values 0 and 1.

\begin{figure}[t]
    \centering
    \subfigure[Message distribution under watermark removal attack.]{\label{fig:audio_watermark_attack_distribution_removal}\includegraphics[width=0.215\textwidth]{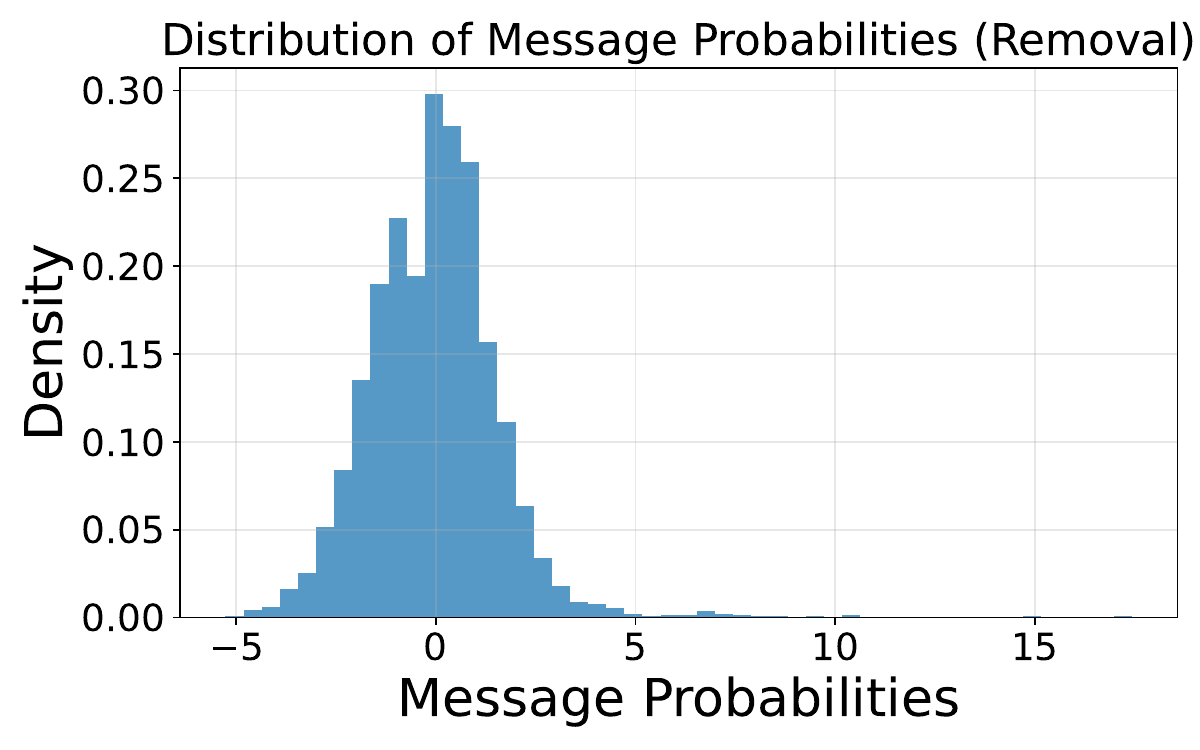}}
    \hspace{0.002\textwidth}
    \subfigure[Message distribution under watermark creation attack.]{\label{fig:audio_watermark_attack_distribution_creation}\includegraphics[width=0.215\textwidth]{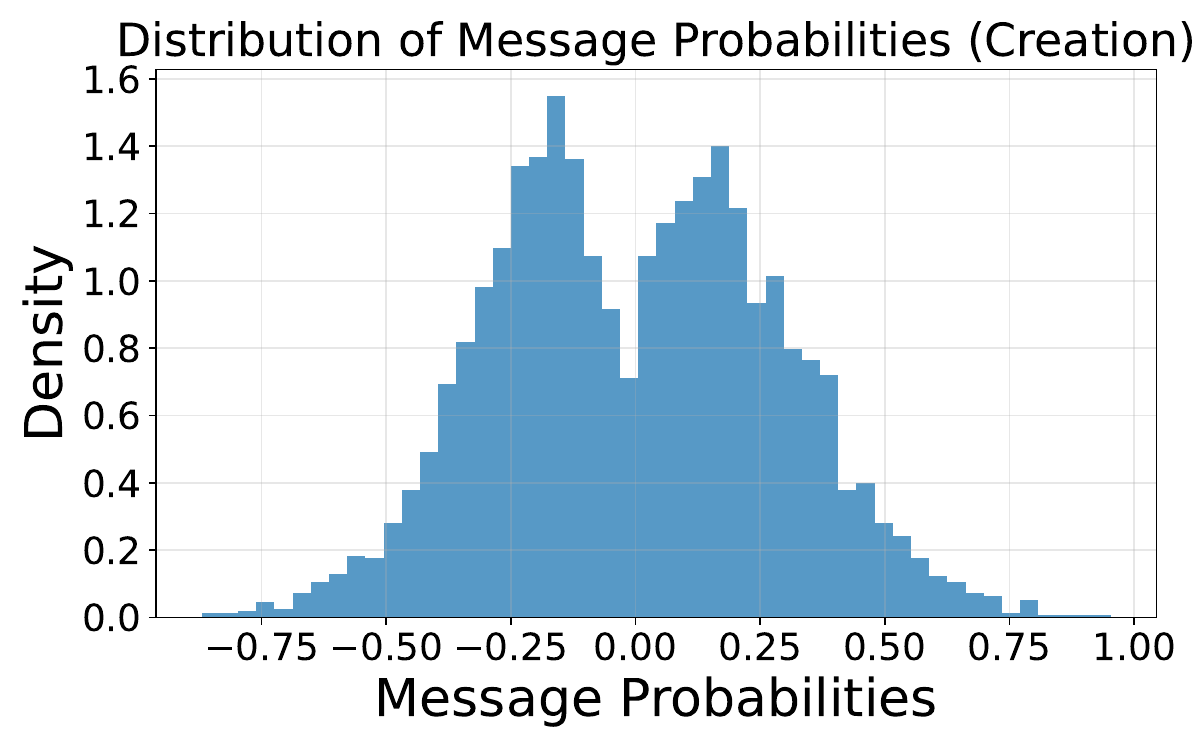}}
    \vspace{-3mm}
    \caption{Distribution of message probabilities under attacks (Timbre). We use the AudioMarkBench to perform the attacks.}
    \label{fig:audio_watermark_attack_distribution}
    \vspace{-3pt}
\end{figure}

\noindent\textbf{Audio Watermark Attack Distribution.} For existing watermark attack methods~\cite{liu2024audiomarkbench}, we observe that the distributions of message probabilities deviate significantly from the benign data distributions. Figure~\ref{fig:audio_watermark_attack_distribution_removal} illustrates the distribution of message probabilities under watermark removal attacks. Compared to Figure~\ref{fig:timbre_normal_distribution_a}, the message distribution under attack differs in the range of message probabilities along the x-axis, even though it still exhibits a unimodal normal distribution. Figure~\ref{fig:audio_watermark_attack_distribution_creation} presents the distribution of watermark creation attacks. Here, most message probability values cluster around the threshold value of $0$. In contrast to the clear bimodal normal distribution shown in Figure~\ref{fig:timbre_normal_distribution_b}, the distribution resulting from the attack visually diverges from the benign data distribution.



\subsection{Threat Model}

The attackers perform watermark attacks to either claim new copyright ownership or remove the original copyright. We assume that: 1) attackers have no access to the data used by defenders to fit the distribution, nor to the audio dataset used to train the watermarking model;
2) they have no access to the ground-truth watermark message embedded in the target audio; but they can get the message probabilities from the watermark decoder, allowing them to perform gradient-based attacks;
3) they do not possess the complete watermarking model, but they can embed watermarks into a small set of clean audio samples;
4) they are aware that the decoded message probabilities output by the watermark decoder follow a normal distribution, but they do not know the corresponding mean and standard deviation. 

The defenders are the attack detectors, who identify whether audio has been tampered with. We assume that: 1) defenders have access to a large number of ground-truth audio samples, which are used to fit a normal distribution and estimate the mean and variance via maximum likelihood estimation;
2) the watermarking model is publicly available through an online platform for commercial use, with a limit on the number of watermarked audio samples that each user can generate per day.

\section{Methodology}\label{sec:design}

\begin{figure}[t]
    \centering
    \includegraphics[width=\linewidth]{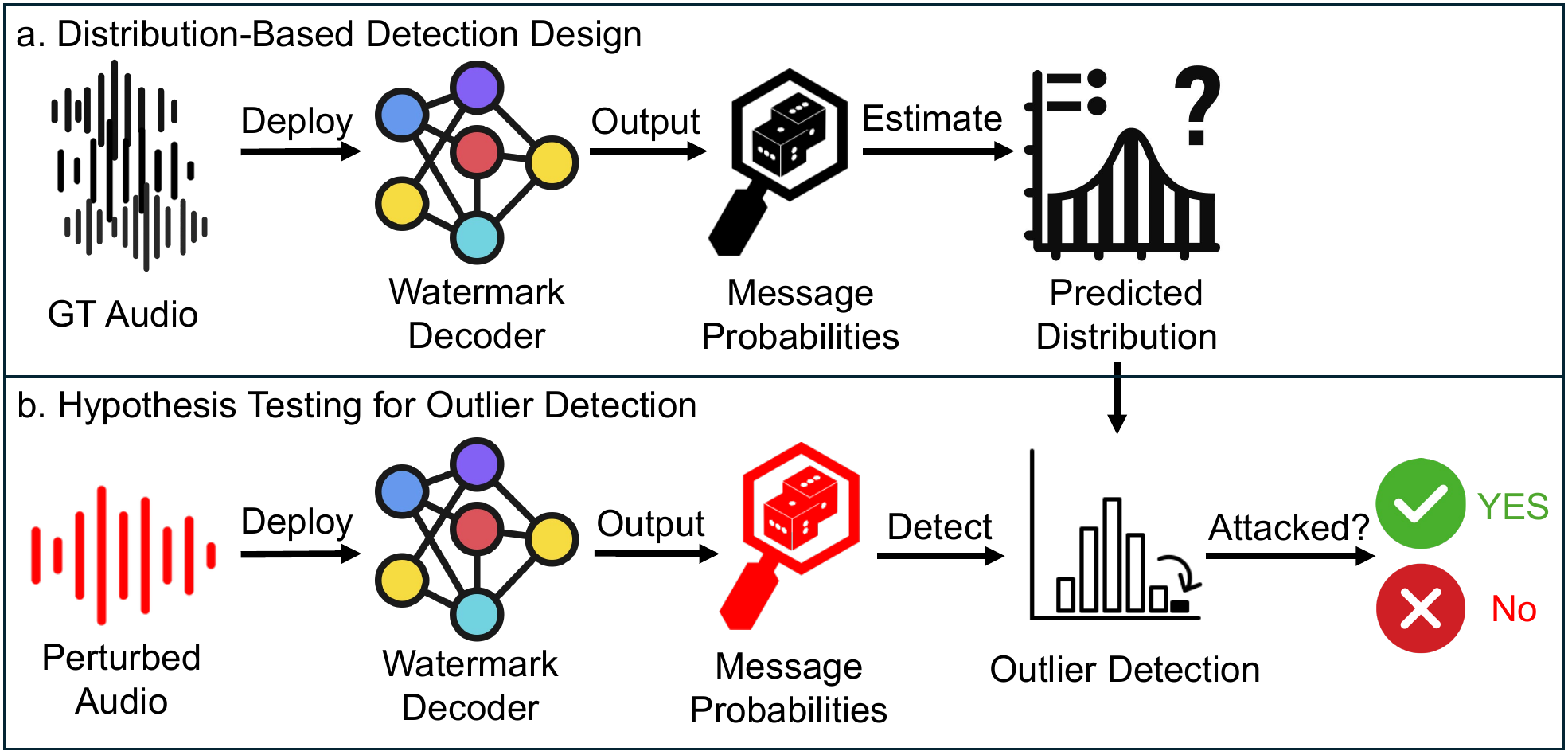}
    \vspace{-6mm}
    \caption{Audio watermark attack detection. The defender uses ground truth (GT) audio (watermarked and clean) to estimate the reference distribution (top) and applies outlier detection with respect to this distribution to determine whether an audio sample has been attacked (bottom).}
    \vspace{-15pt}
    \label{fig:defense}
\end{figure}

In this section, we first introduce a detection method based on outlier detection to identify if the given audio sample has been attacked. Second, we design our adaptive attack, which aims to achieve a successful attack while preserving perceptual quality. 
Meanwhile, we describe the adaptive optimization process for the three attack types.

\subsection{Audio Watermark Attack Detection}\label{sec:detection}





As demonstrated in Section~\ref{subsec:message_probaility_distribution}, the distribution of message probabilities in perturbed audio exhibits significant deviations from that of benign audio, thereby providing the foundation for our proposed detection mechanism to effectively distinguish between the two.
Figure \ref{fig:defense} illustrates the audio watermark attack detection method, which introduces a two-step approach.

\noindent\textbf{Distribution-Based Detection Design.} The key intuition is that message probabilities from clean and watermarked audio follow different statistical patterns. Building on this, we propose that the defenders can collect ground-truth audio, extract message probabilities with the watermark decoder, and apply maximum likelihood estimation to model their distributions. As shown in Figure \ref{fig:normal_distribution_findings}, this process yields three normal distributions: one derived from message probabilities of clean audio and two from those of watermarked audio.


Specifically, given $n$ audio samples, each producing a probability vector of size $1\times N$ message probabilities, the complete set of message probabilities is denoted as:  
$$
\mathbf{p} =
\begin{bmatrix}
p_{11} & p_{12} & \cdots & p_{1N} \\
p_{21} & p_{22} & \cdots & p_{2N} \\
\vdots & \vdots & \ddots & \vdots \\
p_{n1} & p_{n2} & \cdots & p_{nN}
\end{bmatrix}
\in \mathbb{R}^{n \times N}
.$$ 
The mean $\mu$ and standard deviation $\sigma$ of the normal distribution estimated via maximum likelihood are:
\begin{equation}
\begin{aligned}
    & \mu = \frac{1}{nN} \sum_{i=1}^{n} \sum_{j=1}^{N} p_{ij},
    & \sigma^2 = \frac{1}{nN} \sum_{i=1}^{n} \sum_{j=1}^{N} \left( p_{ij} - \mu \right)^{2}.
    \label{eq:maximum_likelihood}
\end{aligned}
\end{equation}
Finally, the defender obtains the corresponding means $\mu$ and standard deviations $\sigma$, which are used to detect outliers in suspicious audio and determine whether it has been attacked.


\noindent\textbf{Hypothesis Testing for Outlier Detection.} We formalize detection as an outlier detection problem under a statistical hypothesis testing framework. Given a potentially perturbed audio sample, the defender uses the watermark decoder to extract its message probability set $\{p_i\}$, where each element is assumed to follow a normal distribution with mean $\mu$ and standard deviation $\sigma$. We then test the null hypothesis $H_0: p_i \sim \mathcal{N}(\mu,\sigma^2)$ against the alternative $H_1: p_i \not\sim \mathcal{N}(\mu,\sigma^2)$. If $H_0$ is rejected, the audio sample is classified as attacked.


To test the hypothesis, we first compute the z-score for each extracted message probability:
\begin{equation}
    z_i = \frac{p_i - \mu}{\sigma}.
\end{equation}

Under the null hypothesis, $z_i$ follows the standard normal distribution, and the corresponding two-tailed p-value is given by $P_i = 2\,(1 - \Phi(|z_i|))$, where $\Phi$ denotes the cumulative distribution function of the standard normal distribution. The decision rule is to reject $H_0$ if $P_i < \alpha$, with $\alpha$ as the predefined significance level. An audio sample is deemed being ``attacked" if any extracted message probability $p_i$ rejects $H_0$.

\subsection{Adaptive Attack Design}~\label{sec:adaptive_attack_design}
We propose an adaptive attack designed to bypass message-probability-distribution–based detection. First, the attacker prepares for the attack by estimating the defender's distribution. Second, the attacker performs the watermark attack, which contains two stages: (1) modify the original audio to bypass the defense and achieve a successful attack, and (2) improve the quality of the perturbed audio, while maintaining that the decoded message probabilities remain within an acceptable range.




\subsubsection{Attack Preparation: Estimate Defender's Distribution}~\label{sec:normal_distribution_estimation}
\begin{figure}[tp]
    \centering
    \includegraphics[width=\linewidth]{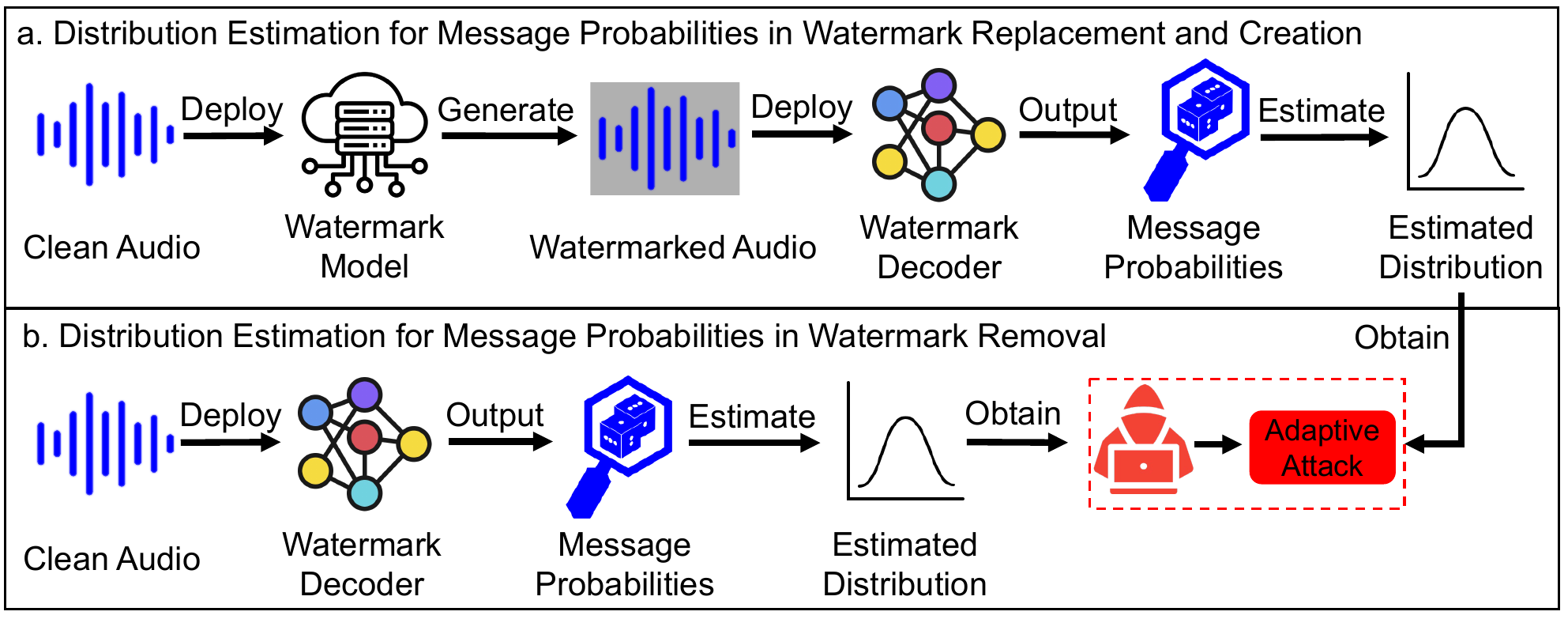}
    \vspace{-6mm}
    \caption{The distribution estimation by the attacker. (a) Watermark replacement and creation: The attacker uses a small set of clean audio samples to generate watermarked audio samples, which are then used to estimate the distribution. (b) Watermark removal: The attacker directly uses the clean audio samples to estimate the distribution.}
    \vspace{-10pt}
    \label{fig:estimate_distribution}
\end{figure}
\begin{figure*}[t]
    \centering
    \includegraphics[width=0.9\linewidth]{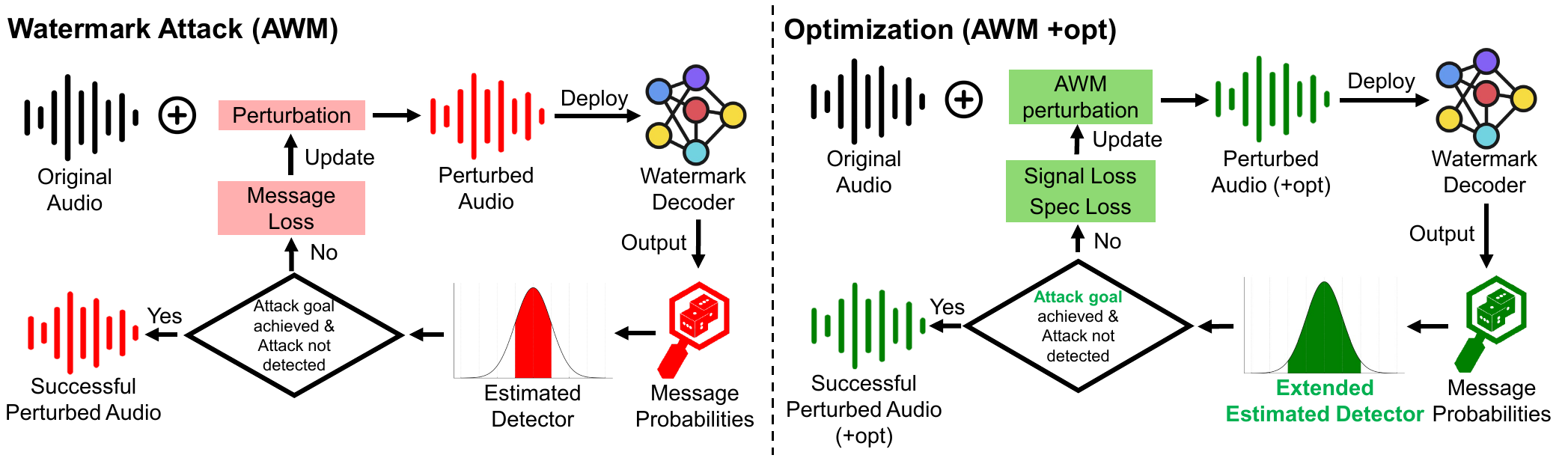}
    \vspace{-10pt}
    \caption{The design of \ours generator. The audio watermark attack step (left) ensures the success of the watermark attack, while the audio quality optimization step (right) focuses on improving audio quality.}
    \vspace{-10pt}
    \label{fig:design}
\end{figure*}
The adversary's goal is to estimate the mean $\mu_{est}$ and the standard deviation $\sigma_{est}$ of the decoded message probabilities. These parameters are categorized into two groups: watermark replacement and creation, and watermark removal. The process is illustrated in Figure \ref{fig:estimate_distribution}.

\noindent\textbf{Parameter Estimation for Watermark Replacement and Creation.} The attackers first select several clean audio samples $s_c$ from a small dataset, which are used to query the watermark model $Enc(\cdot)$ and generate new watermarked audio samples $s_w$. The watermark decoder $Dec(\cdot)$ is deployed to extract message probabilities, which are subsequently used to estimate the parameters of the normal distribution. Since the distribution of watermarked message probabilities is bimodal (as shown in Figure~\ref{fig:timbre_normal_distribution_b} and~\ref{fig:audioseal_normal_distribution_b}), the attacker can obtain two distributions. The estimated mean and standard deviation are as follows:
\begin{equation}
\begin{aligned}
    & \mu_{est}^0, \sigma^0_{est}=T^0(Dec(Enc(s_c))), \\
    & \mu_{est}^1, \sigma^1_{est}=T^1(Dec(Enc(s_c))),
    \label{eq:estimate_distribution_forgery}
\end{aligned}
\end{equation}
where $\mu_{est}^0$ and $\sigma_{est}^0$ represent the estimated mean and standard deviation of decoded message probabilities predicted as 0, and $\mu_{est}^1$ and $\sigma_{est}^1$ correspond to those predicted as 1. $T(\cdot)$ denotes the method used to estimate the parameters of the normal distribution, which depends on the attacker’s prior knowledge and the amount of available data. Common choices include Bayesian inference~\cite{murphy2007conjugate} and maximum likelihood estimation. We use $T^0(\cdot)$ and $T^1(\cdot)$ 
to denote the estimation procedures for the distributions corresponding to bits 0 and 1, respectively.

\noindent\textbf{Parameter Estimation for Watermark Removal.} The attackers estimate the distribution directly using clean audio samples $s_c$. They use the watermark decoder to output message probabilities, which are used for distribution estimation. Since the estimated distribution is unimodal (as shown in Figure~\ref{fig:timbre_normal_distribution_a} and~\ref{fig:audioseal_normal_distribution_a}), the parameters are defined as follows:
\begin{equation}
    \mu_{est}^c, \sigma^c_{est}=T^c(Dec(s_c)),
    \label{eq:estimate_distribution_removal}
\end{equation}
where $\mu_{est}^c$ and $\sigma_{est}^c$ are the estimated mean and standard deviation of the clean message probabilities, and $T^c$ represents the estimation approach applied to these values.

\subsubsection{\textbf{Audio Watermark Attack (\ours)}}~\label{sec:watermark_attack} After obtaining the estimated distribution, the attacker then adds a small adversarial perturbation to the original audio $s$, aiming to deceive the decoder into outputting incorrect binary messages while bypassing the detection strategy. Figure \ref{fig:design}-left illustrates the watermark attack step. The original audio is clean audio in watermark creation attack, and it can be watermarked audio in watermark replacement or watermark removal attack. First, the perturbation is initialized by a fraction of the original audio signal. Next, the attacker adds the perturbation to original audio and obtains a perturbed audio, which is passed through the watermark decoder to obtain message probabilities and subsequently queried by the estimated detector. 
If the attack goal is achieved and the attack is not detected, the attacker obtains the successful perturbed audio. Otherwise, the attacker further optimizes the perturbation by message loss. 
The message loss $\mathcal{L}_{msg}$ modifies the perturbed message probabilities to match the target message probabilities $p_{t}$:
\begin{equation}
    \mathcal{L}_{msg}=\lVert Dec(s_{att})-p_{t})\rVert_2^2.
    \label{eq:msg_loss}
\end{equation}
Let $s_{att}$ denote the perturbed audio, this loss enforces the decoded message to closely match the target message. 

Different from the prior attacks, our loss optimization step follows a strict bit-to-bit optimization design (detailed in Algorithm~\ref{adaptive_optimization_forgery}). This algorithm uses the estimated detector knowledge to ensure that the perturbed audio exhibits a distribution similar to that of benign audio, with high confidence. Importantly, our algorithm constrains the message probabilities within the normal value range. 
Through iterative updates and perturbation optimization, the perturbed message probabilities are gradually adjusted to fall within a strict range, ensuring that they are classified as non-outliers and yielding the final adversarial perturbation.
Our results show that constraining the message probabilities of the perturbed audio to the interval $[\mu_{est} - \sigma_{est}, \mu_{est} + \sigma_{est}]$ improves the attack success rate.

Besides the message loss, we also formulate the signal loss and mel loss to minimize the quality degradation from the attack. Specifically, the signal loss controls the audio quality at the signal level:
\begin{equation}
    \mathcal{L}_{signal}=\frac{1}{n}\sum_{i=1}^{n}|s_{att}-s|.
    \label{eq:wav_loss}
\end{equation}
The Mel-Spectrogram loss $\mathcal{L}_{mel}$ maintains the audio quality at the Mel-Spectrogram level:
\begin{equation}
    \mathcal{L}_{mel}=\lVert Mel(s_{att})-Mel({s})\rVert_2^2.
    \label{eq:mel_loss}
\end{equation}
The total loss in the attack step is:
\begin{equation}
    \mathcal{L}=\lambda_{1}\mathcal{L}_{signal}+\lambda_{2}\mathcal{L}_{mel}+\lambda_{3}\mathcal{L}_{msg}+\lambda_{4}\mathcal{L}_{other},
    \label{eq:total_loss}
\end{equation}
where the $\mathcal{L}_{other}$ depends on the specific watermarking method. For example, AudioSeal~\cite{sanroman2024proactive} includes a localization loss used for indicating the probability of the audio being watermarked. In the \ours attack process, the parameter $\lambda_{msg}$ is assigned a relatively high value.

\subsubsection{Audio Quality Optimization (\ours +opt)}~\label{sec:audio_quality_optimization}  Since audio watermark attacks prioritize message loss optimization, audio quality may be adversely impacted. This step aims to improve audio quality while maintaining a successful attack. To achieve this, we make three adaptations.  

Figure~\ref{fig:design}-right illustrates the three steps. First, the attacker takes the perturbation obtained in the earlier stage as the initial perturbation and uses the watermark decoder to extract the corresponding message probabilities. These probabilities are then fed into the estimated distribution to expand the allowable range, which we suggest setting to $[\mu_{est} - 2\sigma_{est}, \mu_{est} + 2\sigma_{est}]$. This step is crucial for trading off audio quality against attack success rate. In the original \ours setting, the attacker enforces a strict constraint to guarantee attack success, but this rigidity limits the flexibility of the perturbation and makes it difficult to preserve audio quality. In this design, we relax the constraint by expanding the permissible range for message probability optimization, thereby providing more space for perturbation adjustment and achieving a more balanced attack that accounts for both perceptual quality and attack success rate.


Second, in addition to extending the acceptable range, we also update the attack objective by enforcing optimization to proceed for a fixed number of epochs. This ensures that the perturbation is fully optimized rather than stopping at a boundary case. 

Third, we modify the optimization loss by replacing the Mel-spectrogram loss with a standard spectrogram loss and applying a softmax function to the spectrogram. Inspired by AudioSeal~\cite{sanroman2024proactive}, this softmax adaptation better preserves loudness and perceptual similarity between the two audio signals. The resulting softmax-based spectrogram loss, denoted as $\mathcal{L}_{\text{spec}}$, is defined as:
\begin{equation}
    \mathcal{L}_{spec}=\frac{1}{n}\sum_{i=1}^{n}|Softmax(S_{att})-Softmax(S_s)|,
    \label{eq:spec_loss}
\end{equation}
where  $S_{att}$ and $S_s$ are the spectrograms of the attacked audio and original audio. In the optimization (+opt) process, the parameters $\lambda_{1}$ and $\lambda_{2}$ are assigned a relatively high value.

\begin{algorithm}[t]
    \renewcommand{\algorithmicrequire}{\textbf{Input:}}
    \renewcommand{\algorithmicensure}{\textbf{Output:}}
    \caption{Adaptive Attack in Watermark Replacement}
    \label{adaptive_optimization_forgery}
    \begin{algorithmic}[1]
        \Require Watermarked audio $s_w$, target message probabilities $p_{t}$, scale factor $r$, supremum and infimum of thresholds for decoded message 0 and 1 $\tau_{sup}^{0}$ and $\tau_{infi}^{0}$, $\tau_{sup}^{1}$ and $\tau_{infi}^{1}$, list of different message probabilities $msg_{diff}$ 
        \Ensure perturbed watermarked audio $\hat{s_w}$ 
 
        \State $\delta = s_w \times r$, $p_w = Dec(s_w)$, $\hat{s_w} = s_w + \delta$

        \For {$index=1,2,\cdots,(len(p_{t})-1)$}
            \If {$index$ not in $msg_{diff}$}
                \State $p_{t}[index] = p_w[index]$
            \EndIf
        \EndFor

        \For {$i=1,2,\cdots,Iter$}
            \State $\delta = Attack(s_w,\hat{s_w}, Dec(\hat{s_w}), p_{t}, \delta)$
            \State $\hat{s_w} = s_w+\delta$
            \State $\hat{p_w} = Dec(\hat{s_w})$
            \If {$acc==1$}
                \For {$index$ in $msg_{diff}$}
                    \If {$\tau_{infi}^{1} < \hat{p_w }[index] < \tau_{sup}^{1}$}
                        \State $p_{t}[index] = \hat{p_w}[index]$
                        \State Remove $msg_{diff}[index]$
                    \EndIf
                    \If {$\tau_{infi}^{0} < \hat{p_w}[index] < \tau_{sup}^{0}$}
                        \State $p_{t}[index] = \hat{p_w}[index]$
                        \State Remove $msg_{diff}[index]$
                    \EndIf
                \EndFor
            \EndIf
            \If {meet the estimated detection $Detection(\hat{p_w})$}
                \State \Return $\hat{s_w}$
            \EndIf
        \EndFor
        \State \Return \textbf{Failed}
    \end{algorithmic}
\end{algorithm}

\subsubsection{Adaptive Attack for Watermark Replacement, Creation, and Removal} During the watermark attack, we prioritize refining the decoded message probabilities that fall outside the estimated normal range. Algorithm \ref{adaptive_optimization_forgery} outlines the adaptive attack process used in watermark replacement. For example, suppose the original watermark bits are ``101100", which are modified to ``111000". 
We define a list, $msg_{diff}$, which contains the indices of binary message bits differing between the original watermarked audio $s_{w}$ and the perturbed watermarked audio $\hat{s_{w}}$\footnote{This notation is defined for the watermark replacement scenario. The general representation of perturbed audio is denoted as $s_{att}$, which refers either to perturbed clean audio (used for creation) or to perturbed watermarked audio (used for replacement or removal).}. In this case, $msg_{diff} = [1, 3]$. For the indices not included in $msg_{diff}$, we assign the original watermark message probabilities $p_w$ to the corresponding target watermark message probabilities $p_{t}$. That is, for indices $[0,2,4,5]$, the $p_{t}$ is equal to the $p_w$. 

This optimization has two advantages: (1) It directs the gradient to focus more on the indices where the binary message bits require modification. In watermark replacement, only the differing bits need to be changed, so bits that already align with the target message do not require further optimization. (2) It ensures that the message probabilities in the non-change bits remain the same, and within the acceptable distribution. 
In certain audio watermarking methods, message probabilities do not directly map to binary message values. In Figure~\ref{fig:audioseal_normal_distribution_b}, probabilities corresponding to binary 1 lie in the range of 0.7–0.8, and those for binary 0 fall between 0.2–0.3. Therefore, it is not appropriate to optimize message probabilities directly to 1 or 0. Instead, they should be adjusted to fall within the range of 0.7–0.8 for a binary message bit of 1, and 0.2–0.3 for a bit of 0. Then, we optimize the perturbation $\delta$ by Function $Attack(\cdot)$ (Line 6) to generate the perturbed watermarked audio $\hat{s_w}$. The $Attack(\cdot)$ calculates the loss in Equation~(\ref{eq:total_loss}) and use the gradient to update the perturbation.
The initial perturbation is scaled based on the original watermarked audio $s_w$ to ensure that the decoded message probabilities of the perturbed audio $\hat{s_w}$ closely resemble those of the original. In the attack justification step, we define the supremum and infimum thresholds for the watermark decoder outputs corresponding to binary message bits. Specifically, $\tau_{sup}^{0}$ and $\tau_{infi}^{0}$ represent the thresholds for bit 0, while $\tau_{sup}^{1}$ and $\tau_{infi}^{1}$ correspond to bit 1:
\begin{equation}
\begin{aligned}
    & (\mu_{est}^{0}-\sigma_{est}^0)\leq\tau_{infi}^{0}<\tau_{sup}^{0}\leq(\mu_{est}^0+\sigma_{est}^0), \\
    & (\mu_{est}^{1}-\sigma_{est}^1)\leq\tau_{infi}^{1}<\tau_{sup}^{1}\leq(\mu_{est}^1+\sigma_{est}^1).
    \label{eq:range_threshold}
\end{aligned}
\end{equation}

Once the message probability of the perturbed watermarked audio at a given index falls within the specified threshold range, it is assigned to the target message probability $p_{t}$, and the index is removed from the list $msg_{diff}$. Optimization then proceeds with the remaining message probabilities in the list. To ensure that all perturbed message probabilities remain within the estimated normal range, the attacker simulates the defender's role by performing outlier detection.

For watermark creation attack, the adaptive attack process is the same as that of watermark replacement. The difference is that, in watermark creation, $msg_{diff}$ includes all message indices, which is equal to the full binary message length. Watermark removal attack is an untargeted attack, it is not necessary to achieve an accuracy of exactly 0, any value below 1 is sufficient. Using a lower accuracy threshold requires more iteration steps to optimize the perturbation $\delta$. We recommend using an accuracy threshold around 0.5. 

\section{Evaluation}\label{sec:eva}


\subsection{Experimental Setup} \label{sec:experiment_setup}

\noindent\textbf{Datasets.}
We use three public datasets for our experiments. The first dataset is the LibriSpeech~\cite{panayotov2015librispeech}. We select the small-sized subset, which has 6.3G audios, and covers 100.6 hours of audio data spoken by 251 speakers. The second dataset is obtained from AudioMarkData~\cite{liu2024audiomarkbench}, which is built based on the Common Voice dataset~\cite{ardila2019common}. It contains 20,000 audio samples, each with a duration of 5 seconds. The third dataset is GigaSpeech~\cite{chen2021gigaspeech}, which includes audio from audiobooks, podcasts, and YouTube. We use the XS subset, which contains a total of 10 hours of audio samples.

\noindent\textbf{Audio Watermark Methods.}
We select two state-of-the-art audio watermarking methods: Timbre~\cite{timbrewatermarking-ndss2024} and AudioSeal~\cite{sanroman2024proactive}. We fix the binary message length to 16 bits for all experiments.

\noindent\textbf{Evaluation Metrics.}
First, we introduce the \textbf{Detection Success Rate (DSR)}, which measures the ability to identify outliers in the decoded message probabilities. In this experiment, the defender defines the predefined significance level $\alpha$ as 0.003. Second, we use the \textbf{False Acceptance Rate (FAR)}, which evaluates the rate that unattacked audio is mistakenly classified as attacked. Third, we employ the \textbf{Attack Success Rate (ASR)} to measure \emph{how successfully the adaptive attackers perform the watermark attack}, which also corresponds to the watermark decoder's accuracy at the binary message level.

For the audio quality metrics, we use the \textbf{Signal-to-noise ratio (SNR)} and \textbf{ViSQOL}~\cite{hines2015visqol}. In the experiment, we use clean audio samples as a baseline.


\begin{table}[t]
\centering
\scriptsize
\setlength{\tabcolsep}{4pt}
\renewcommand{\arraystretch}{1.2}
\caption{\textnormal{Detection performance across different datasets, watermark methods, and attack methods. '-' indicates that no successfully perturbed audio is available.}}
\vspace{-2mm}
\scalebox{0.5}{
\begin{tabular}{|c|c|c|c|c|c|c|c|c|c|c|c|}
\hline
\multirow{2}{*}{\textbf{Attack Type}} & \multirow{2}{*}{\textbf{\shortstack{Watermark \\ Method}}} & \multirow{2}{*}{\textbf{Attack Method}} & \multicolumn{3}{c|}{\textbf{Librispeech}} & \multicolumn{3}{c|}{\textbf{Audiomark}} & \multicolumn{3}{c|}{\textbf{Gigaspeech}} \\
\cline{4-12}
 & & & \textbf{DSR (\%)} & \textbf{FAR (\%)} & \textbf{F1 (\%)} & \textbf{DSR (\%)} & \textbf{FAR (\%)} & \textbf{F1 (\%)} & \textbf{DSR (\%)} & \textbf{FAR (\%)} & \textbf{F1 (\%)} \\
\hline
\multirow{16}{*}{\shortstack{Watermark \\ Replacement}}
& \multirow{8}{*}{AudioSeal} & AudioMarkBench  & 97.71 & 4.20 & 96.79 & 100.00 & 5.50 & 97.32 & 100.00 & 6.00 & 96.94 \\
& & AudioM. (+LP)  & 93.94 & 2.27 & 95.75 & 93.98 & 4.82 & 89.66 & 96.05 & 8.45 & 94.19 \\
& & AudioM. (+AS)  & 97.96 & 4.08 & 96.97 & 100.00 & 5.23 & 97.45 & 99.61 & 5.85 & 97.53 \\
& & AudioM. (+GN)  & 98.03 & 1.97 & 98.03 & 100.00 & 10.81 & 94.87 & 98.11 & 7.14 & 98.85 \\
& & AudioM. (+MP3)  & 98.47 & 4.20 & 97.08 & 100.00 & 7.23 & 96.51 & 96.93 & 3.40 & 96.93 \\
& & AudioM. (+HP)  & 86.41 & 3.88 & 90.82 & 96.43 & 5.36 & 95.58 & 68.75 & 4.76 & 79.49 \\
& & Ours  & \cellcolor{red!25} 3.44 & \cellcolor{red!25}4.20 & \cellcolor{red!25}6.40 & \cellcolor{red!25}8.50 & \cellcolor{red!25}5.50 & \cellcolor{red!25}15.00 & \cellcolor{red!25}11.00 & \cellcolor{red!25}6.00 & \cellcolor{red!25}18.81 \\
& & Ours (+opt) & \cellcolor{green!25}5.34 & \cellcolor{green!25}4.20 & \cellcolor{green!25}9.75 & \cellcolor{green!25}8.00 & \cellcolor{green!25}5.50 & \cellcolor{green!25}14.10 & \cellcolor{green!25}17.67 & \cellcolor{green!25}6.00 & \cellcolor{green!25}28.58 \\
\cline{2-12}
& \multirow{8}{*}{Timbre} & AudioMarkBench & 100.00 & 2.67 & 98.68 & 100.00 & 6.50 & 96.85 & 100.00 & 8.33 & 96.00 \\
& & AudioM. (+LP)  & 100.00 & 0.00 & 100.00 & 100.00 & 9.09 & 95.65 & 100.00 & 0.00 & 100.00 \\
& & AudioM. (+AS)  & 100.00 & 2.82 & 98.61 & 100.00 & 4.48 & 97.81 & 100.00 & 7.64 & 96.49 \\
& & AudioM. (+GN)  & - & - & - & 100.00 & 0.00 & 100.00 & - & - & - \\
& & AudioM. (+MP3)  & 100.00 & 0.00 & 100.00 & 100.00 & 0.00 & 100.00 & - & - & - \\
& & AudioM. (+HP)  & 100.00 & 1.28 & 99.36 & 100.00 & 2.94 & 98.55 & 100.00 & 4.00 & 98.11 \\
& & Ours & \cellcolor{red!25}1.53 & \cellcolor{red!25}2.67 & \cellcolor{red!25}2.93 & \cellcolor{red!25}7.00 & \cellcolor{red!25}6.50 & \cellcolor{red!25}12.33 & \cellcolor{red!25}6.67 & \cellcolor{red!25}8.33 & \cellcolor{red!25}11.58 \\
& & Ours (+opt) & \cellcolor{green!25}1.91 & \cellcolor{green!25}2.67 & \cellcolor{green!25}3.64 & \cellcolor{green!25}7.50 & \cellcolor{green!25}6.50 & \cellcolor{green!25}13.17 & \cellcolor{green!25}8.33 & \cellcolor{green!25}8.33 & \cellcolor{green!25}14.29 \\
\hline
\multirow{16}{*}{\shortstack{Watermark \\ Creation}}
& \multirow{8}{*}{AudioSeal} & AudioMarkBench & 100.00 & 4.20 & 97.94 & 100.00 & 5.50 & 97.32 & 100.00 &6.00 & 97.09 \\
& & AudioM. (+LP)  & 100.00 & 4.90 & 97.61 & 100.00 & 2.82 & 98.61 & 100.00 & 9.00 & 95.96 \\
& & AudioM. (+AS)  & 100.00 & 4.50 & 97.79 & 100.00 & 3.97 & 98.05 & 100.00 & 6.76 & 97.18 \\
& & AudioM. (+GN)  & 100.00 & 5.22 & 97.46 & 100.00 & 0.00 & 100.00 & 100.00 & 14.89 & 93.33 \\
& & AudioM. (+MP3)  & 100.00 & 5.45 & 97.35 & 100.00 & 4.76 & 97.67 & 100.00 & 11.76 & 94.74 \\
& & AudioM. (+HP)  & 100.00 & 1.28 & 99.36 & 100.00 & 2.94 & 98.55 & 100.00 & 4.00 & 98.11 \\
& & Ours & \cellcolor{red!25}0.76 & \cellcolor{red!25}4.20 & \cellcolor{red!25}1.45 & \cellcolor{red!25}0.50 & \cellcolor{red!25}5.50 & \cellcolor{red!25}0.94 & \cellcolor{red!25}0.33 & \cellcolor{red!25}6.00 & \cellcolor{red!25}0.63 \\
& & Ours (+opt) & \cellcolor{green!25}1.91 & \cellcolor{green!25}4.20 & \cellcolor{green!25}3.60 & \cellcolor{green!25}2.00 & \cellcolor{green!25}5.50 & \cellcolor{green!25}3.68 & \cellcolor{green!25}13.00 & \cellcolor{green!25}6.00 & \cellcolor{green!25}21.85 \\
\cline{2-12}
& \multirow{8}{*}{Timbre} & AudioMarkBench & 100.00 & 2.67 & 98.68 & 100.00 & 6.50 & 96.85 & 100.00 & 8.33 & 96.00 \\
& & AudioM. (+LP)  & 100.00 & 2.06 & 99.00 & 100.00 & 5.31 & 97.41 & 100.00 & 6.09 & 97.30 \\
& & AudioM. (+AS)  & 100.00 & 2.31 & 98.86 & 100.00 & 6.40 & 96.90 & 100.00 & 8.25 & 96.59 \\
& & AudioM. (+GN)  & 100.00 & 0.00 & 100.00 & 100.00 & 33.33 & 85.71 & 100.00 & 0.00 & 100.00 \\
& & AudioM. (+MP3)  & 100.00 & 3.70 & 98.18 & 100.00 & 3.70 & 98.18 & 100.00 & 4.55 & 97.87 \\
& & AudioM. (+HP)  & 100.00 & 2.60 & 98.71 & 100.00 & 5.71 & 97.22 & 100.00 & 6.53 & 97.25 \\
& & Ours & \cellcolor{red!25}0.00 & \cellcolor{red!25}2.67 & \cellcolor{red!25}0.00 & \cellcolor{red!25}0.00 & \cellcolor{red!25}6.50 & \cellcolor{red!25}0.00 & \cellcolor{red!25}0.00 & \cellcolor{red!25}8.33 & \cellcolor{red!25}0.00 \\
& & Ours (+opt) & \cellcolor{green!25}0.00 & \cellcolor{green!25}2.67 & \cellcolor{green!25}0.00 & \cellcolor{green!25}0.00 & \cellcolor{green!25}6.50 & \cellcolor{green!25}0.00 & \cellcolor{green!25}0.00 & \cellcolor{green!25}8.33 & \cellcolor{green!25}0.00 \\
\hline
\multirow{16}{*}{\shortstack{Watermark \\ Removal}}
& \multirow{8}{*}{AudioSeal} & AudioMarkBench & 92.75 & 5.73 & 93.46 & 100.00 & 3.50 & 98.28 & 99.67 & 5.67 & 97.03 \\
& & AudioM. (+LP)  & 91.41 & 5.86 & 93.62 & 100.00 & 3.30 & 98.38 & 99.64 & 4.57 & 97.98 \\
& & AudioM. (+AS)  & 91.92 & 5.77 & 93.03 & 100.00 & 2.75 & 98.65 & 99.63 & 5.14 & 97.86 \\
& & AudioM. (+GN)  & 89.92 & 5.43 & 92.00 & 100.00 & 2.96 & 98.55 & 99.63 & 4.78 & 97.95 \\
& & AudioM. (+MP3)  & 83.92 & 5.88 & 88.25 & 100.00 & 2.76 & 98.63 & 98.87 & 4.81 & 97.44 \\
& & AudioM. (+HP)  & 83.92 & 5.88 & 88.25 & 100.00 & 3.39 & 98.33 & 98.17 & 5.58 & 96.95 \\
& & Ours & \cellcolor{red!25}0.00 & \cellcolor{red!25}5.73 & \cellcolor{red!25}0.00 & \cellcolor{red!25}0.00 & \cellcolor{red!25}3.50 & \cellcolor{red!25}0.00 & \cellcolor{red!25}0.00 & \cellcolor{red!25}5.67 & \cellcolor{red!25}0.00 \\
& & Ours (+opt) & \cellcolor{green!25}0.00 & \cellcolor{green!25}5.73 & \cellcolor{green!25}0.00 & \cellcolor{green!25}0.00 & \cellcolor{green!25}3.50 & \cellcolor{green!25}0.00 & \cellcolor{green!25}0.00 & \cellcolor{green!25}5.67 & \cellcolor{green!25}0.00 \\
\cline{2-12}
& \multirow{8}{*}{Timbre} & AudioMarkBench & 100.00 & 5.73 & 97.27 & 100.00 & 6.00 & 97.09 & 100.00 & 8.67 & 95.83 \\
& & AudioM. (+LP)  & 100.00 & 5.73 & 97.22 & 100.00 & 6.00 & 97.09 & 100.00 & 8.98 & 96.50 \\
& & AudioM. (+AS)  & 100.00 & 5.73 & 97.22 & 100.00 & 6.00 & 97.09 & 100.00 & 8.98 & 96.50 \\
& & AudioM. (+GN)  & 84.13 & 5.56 & 81.54 & 74.83 & 7.29 & 82.20 & 85.04 & 9.73 & 88.09 \\
& & AudioM. (+MP3)  & 100.00 & 5.73 & 97.22 & 100.00 & 6.00 & 97.09 & 98.67 & 8.98 & 95.80 \\
& & AudioM. (+HP)  & 100.00 & 5.73 & 97.22 & 100.00 & 6.00 & 97.09 & 100.00 & 8.98 & 96.50 \\
& & Ours & \cellcolor{red!25}0.00 & \cellcolor{red!25}5.73 & \cellcolor{red!25}0.00 & \cellcolor{red!25}0.00 & \cellcolor{red!25}6.00 & \cellcolor{red!25}0.00 & \cellcolor{red!25}0.00 & \cellcolor{red!25}8.67 & \cellcolor{red!25}0.00 \\
& & Ours (+opt) & \cellcolor{green!25}0.00 & \cellcolor{green!25}5.73 & \cellcolor{green!25}0.00 & \cellcolor{green!25}0.00 & \cellcolor{green!25}6.00 & \cellcolor{green!25}0.00 & \cellcolor{green!25}0.00 & \cellcolor{green!25}8.67 & \cellcolor{green!25}0.00 \\
\hline
\end{tabular}}
\label{tab:detection_performance}
\end{table}

\begin{figure}[t]
    \centering     
    \includegraphics[width=\linewidth]{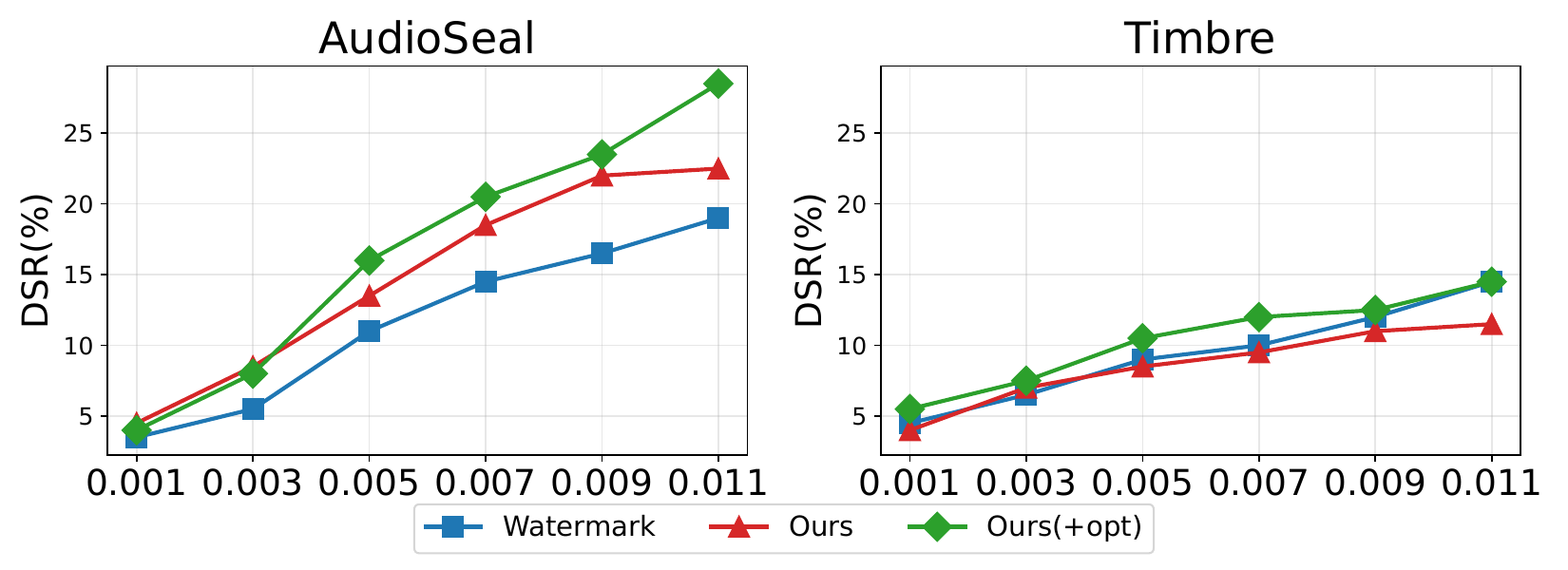}
    \vspace{-7mm}
    \caption {DSR of the watermark replacement attacks under different significance levels $\alpha$.}
    \vspace{-3mm}
    \label{fig:DSR_pvalue}
\end{figure}

\noindent\textbf{Audio Watermark Attack Methods.}
We compare our attack method with the AudioMarkBench~\cite{liu2024audiomarkbench}. AudioMarkBench is a benchmark to evaluate the robustness of audio watermarking against watermark replacement, creation, and removal attacks. We use the default settings if there is no further mention. Our method consists of two attack variants: Ours (\ours) (Section~\ref{sec:watermark_attack}), which includes only the watermark attack step, and Ours (+opt) (\ours+opt) (Section~\ref{sec:audio_quality_optimization}), which includes an additional audio quality optimization step.


\begin{figure}[t]
    \centering
    \subfigure[Message probabilities distribution (AudioSeal)]{
    \includegraphics[width=0.22\textwidth]{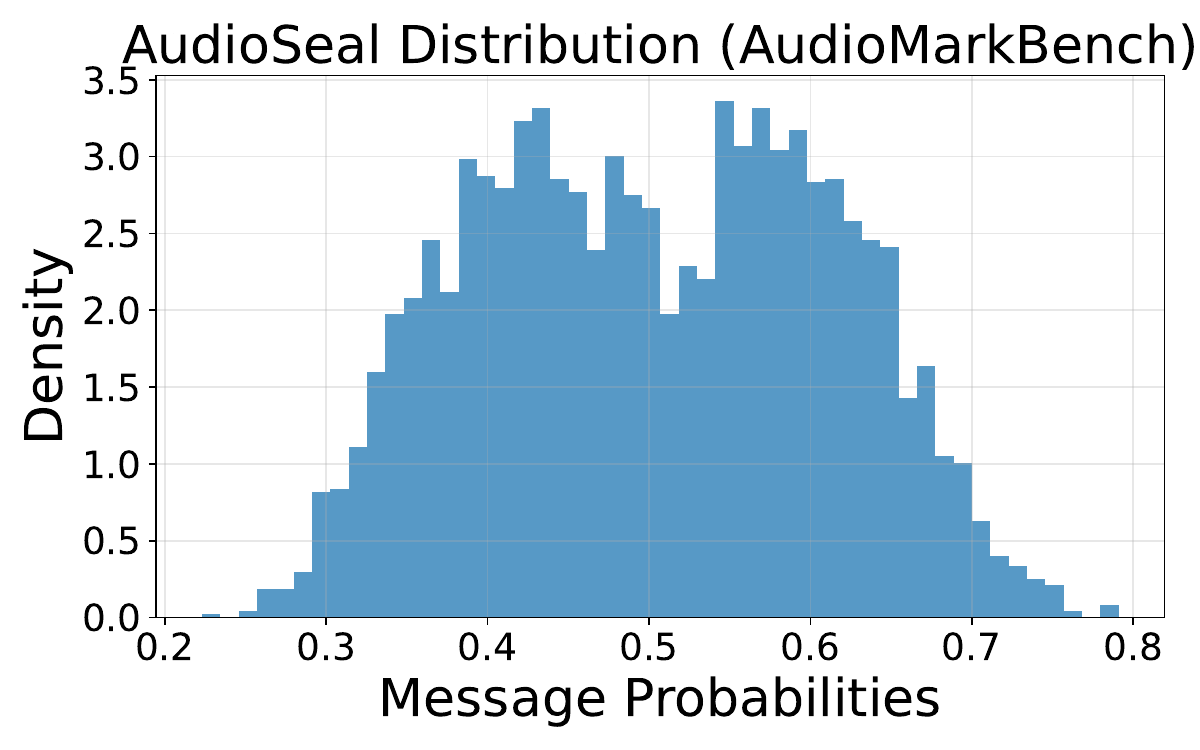}
    \includegraphics[width=0.22\textwidth]{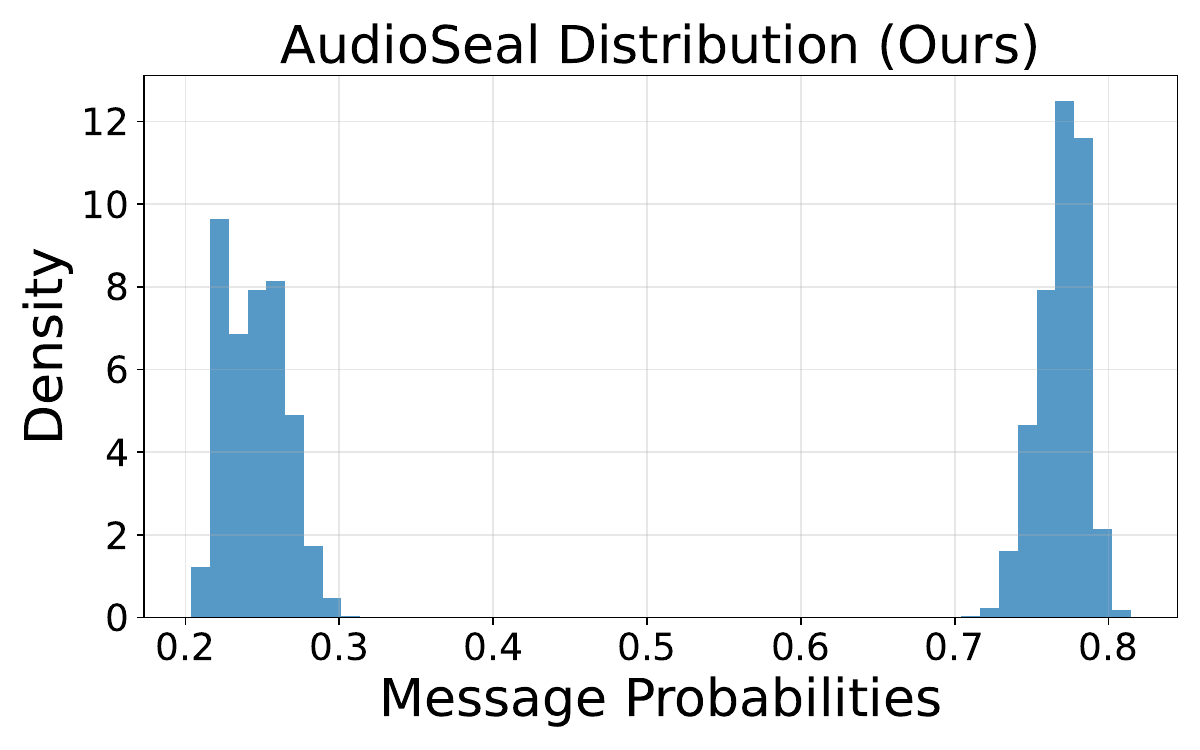}\label{fig:distribution_compare_audioseal}}
    \subfigure[Message probabilities distribution (Timbre)]{
    \includegraphics[width=0.22\textwidth]{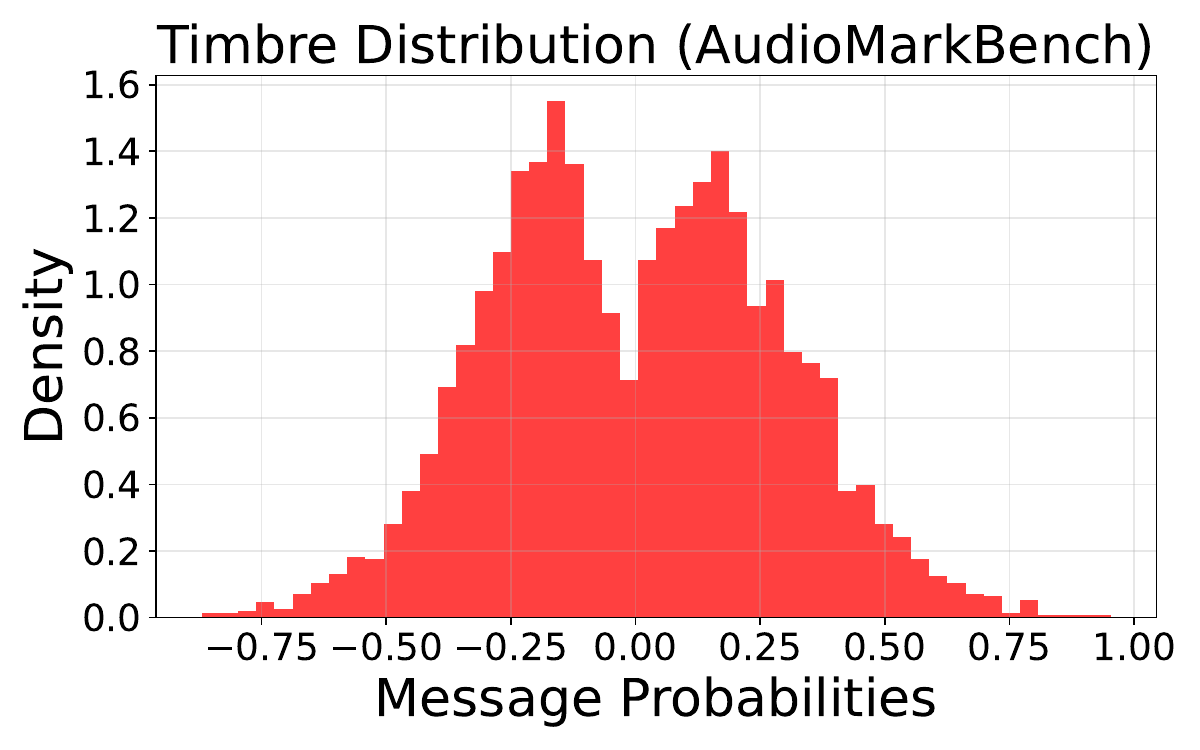}
    \includegraphics[width=0.22\textwidth]{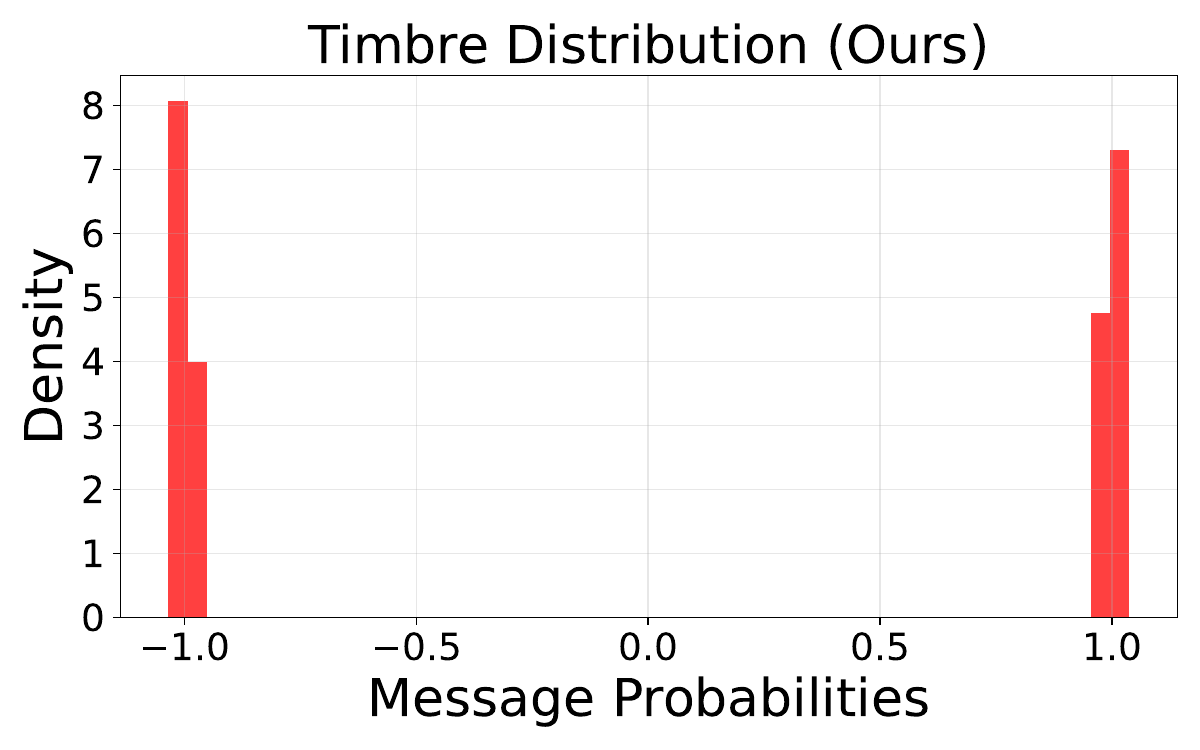}}
    \vspace{-3mm}
    \caption{Message probabilities distribution comparisons between AudioMarkBench and \ours for the watermark creation.}
    \vspace{-13pt}
    \label{fig:distribution_compare}
\end{figure}

\subsection{Attack Detection Performance} \label{subsec:detection_result}
In our experiments, we evaluate our detection method described in Section~\ref{sec:detection} against the attacks provided in AudioMarkBench~\cite{liu2024audiomarkbench}, as well as five additional perturbations applied to the attacked audio: (1) low-pass filtering (LP): fixed at 5000Hz, (2) amplitude scaling (AS): fixed at scale of 0.9, (3) Gaussian noise (GN): fixed at 40 dB, (4) MP3 compression (MP3): fixed at 32kbps, and (5) high-pass filtering (HP): fixed at 250Hz
on AudioSeal and fixed at 500Hz (replacement and removal) and 1000Hz (creation) on Timbre. The detection results are reported in Table~\ref{tab:detection_performance}, where the FAR remains within an acceptable range, with most values around 5\%.

Across all three attack types and eight attack methods, \ours consistently achieves superior evasive performance compared to the baselines. Moreover, the DSR of our watermark attack (Ours) is lower than that of the variant with audio-quality optimization, Ours(+opt). For watermark replacement and creation, most DSR values remain below 10\%, and the best performance is observed in watermark removal, where none of the perturbed audio samples are detected by the defenders. These results suggest that watermark removal is the most effective attack strategy, while watermark replacement is the most challenging among the three evaluated attack types.


\begin{figure}[t]
    \centering
    \includegraphics[width=1\linewidth]{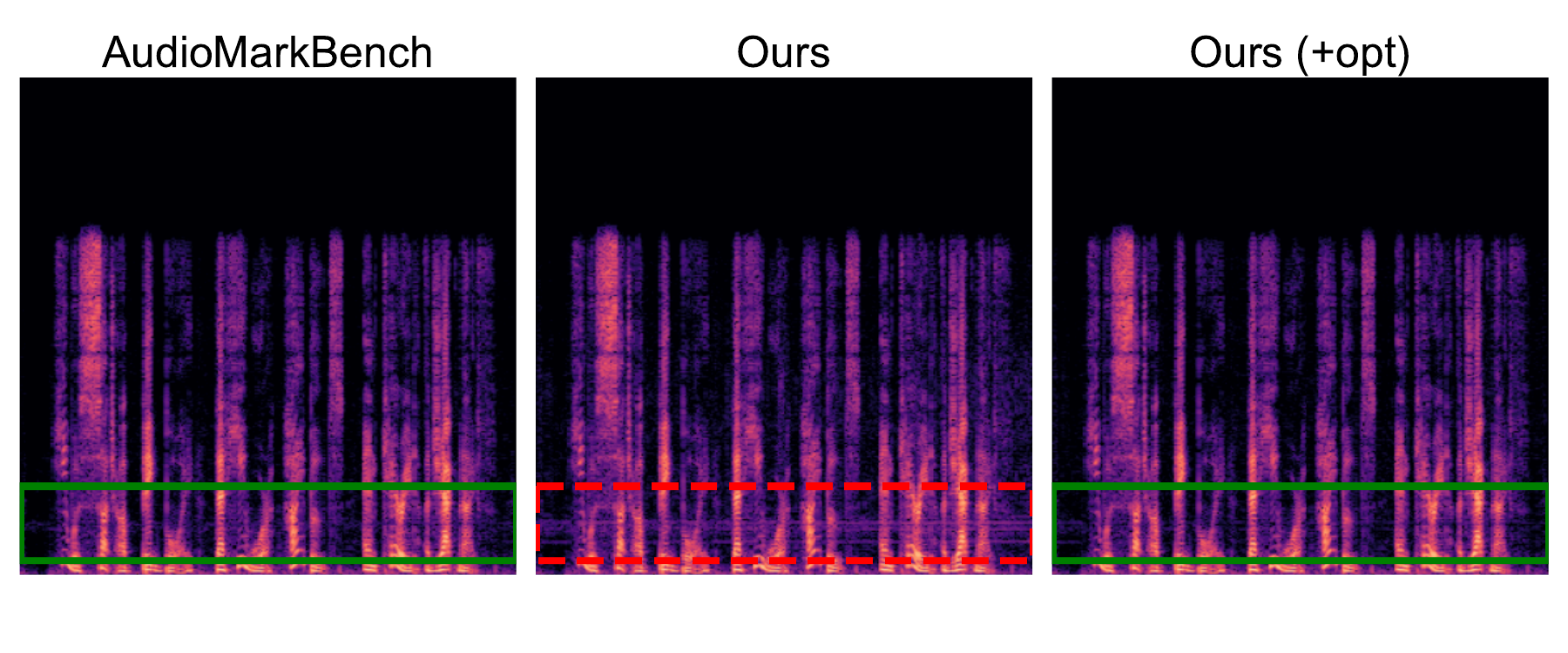}
    \vspace{-5mm}\caption{The spectrograms of the watermark creation in AudioSeal. The dotted red box shows noticeable noise for \ours attack. The green box indicates that \ours (+opt) attack achieves higher audio quality and is visually similar to AudioMarkBench.}
    \vspace{-10pt}
    \label{fig:spec_creation}
\end{figure}

\begin{figure*}[t]
\centering
\includegraphics[width=0.74\linewidth]{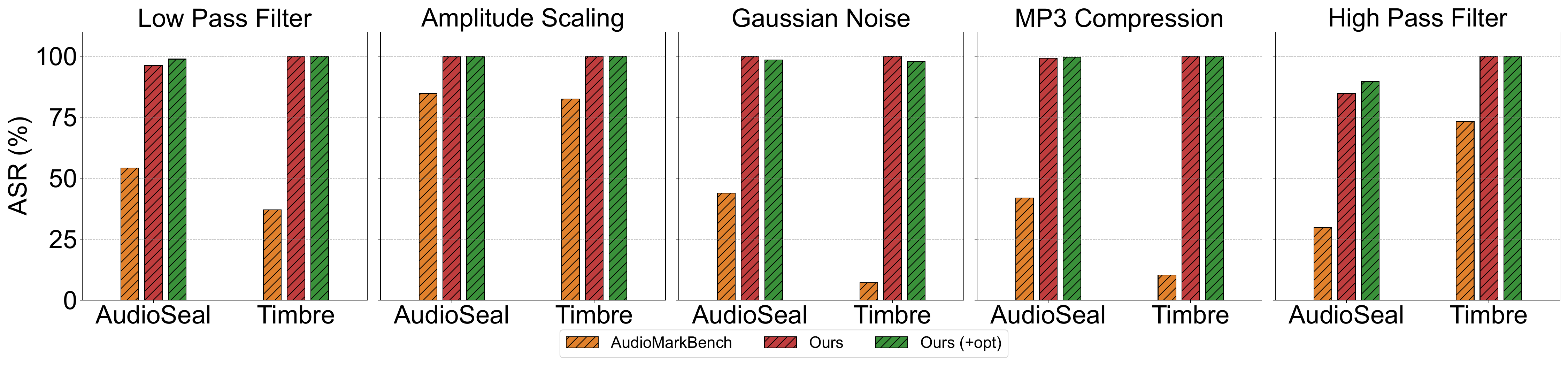}
\vspace{-4mm}
\caption {ASR after applying different no-box perturbations on the watermark creation attacks. It evaluates the ASR results of watermark creation against the no-box perturbations, which uses the Librispeech dataset for validation.}
\vspace{-4mm}
\label{fig:watermark_creation}
\end{figure*}

\subsection{Impact of Different $\alpha$} \label{sec:fig:DSR_pvalue}
As a defender, it is essential to select an appropriate significance level $\alpha$ (as presented in Section~\ref{sec:detection}) for balancing detection and false acceptance. A higher $\alpha$ increases the FAR, while a lower $\alpha$ risks missing detection for the attacked samples. Figure~\ref{fig:DSR_pvalue} shows the DSR under different $\alpha$ values on the Audiomark dataset for the watermark replacement attack. Notably, the DSR of benign watermark audio samples (“Watermark”) is equal to the FAR. We observe that the DSR of the attack ``Ours" is lower than that of the attack variant ``Ours (+opt)", which indicates that improving the audio quality will increase the detection risk.
Since the DSR of the attacked audio closely matches that of ground-truth watermarked audio, \ours attack remains highly stealthy, making it difficult for defenders to detect regardless of the data size or significance level used.

\subsection{Distribution Analysis of \ours} \label{sec:distribution}

We randomly select some audio samples from the Librispeech dataset, and use the AudioMarkBench and our attack method to generate the distribution of message probabilities. Figure \ref{fig:distribution_compare} illustrates the normal distributions for both the AudioSeal and Timbre. In AudioSeal, the predefined threshold $\theta$ for converting probabilities into binary message values is 0.5, and in Timbre, it is 0. The results from AudioMarkBench show that most message probabilities cluster around the predefined threshold, and the overall distribution tends to exhibit a unimodal shape. Therefore, the detection method is able to identify that the audio has been tampered with. In contrast, for our method, the resulting ``normal” distribution becomes bimodal, with a shape similar to that shown in Figure \ref{fig:timbre_normal_distribution_b} and \ref{fig:audioseal_normal_distribution_b}. As a result, our attack successfully bypasses this detection strategy.



\subsection{Perturbation Visualization} \label{sec:visual}

To analyze the attack visually, we generate spectrograms of the audio samples. Figure \ref{fig:spec_creation} presents the visualization using dB-scaled spectrograms of the watermark creation attack. In the spectrograms produced by AudioMarkBench and \ours's default attack method, we can clearly observe some noticeable noise (horizontal lines), which is highlighted with dotted red boxes. Since watermark attacks target only the decoder, some noise may not be optimized well. Comparing the attack methods \ours (Ours) and \ours+opt (Ours +opt), we observe that the audio quality improves and some of the noticeable noise is effectively reduced through optimization. Besides, \ours+opt is more visually similar to AudioMarkBench (green boxes).



\subsection{Robustness Against Perturbations} \label{sec:robustness}

Robustness of the attack in watermarking refers to whether the watermark remains detectable after applying no-box perturbations to the watermarked audio. Watermark creation attack, which aims to modify the binary message bits to the specific target binary message bits while bypassing the defender’s detection across all bits, requires complex and robust considerations. Therefore, we add no-box perturbations to the perturbed watermarked audio and observe if the forged watermark can be detected. Figure \ref{fig:watermark_creation} evaluates the robustness of the watermark creation attack against no-box perturbations. The results show that \ours attack is more resilient against the no-box perturbations, with most ASR achieving around 100\%. For comparing the ASR between \ours and \ours (+opt), \ours attack without optimization demonstrates slightly higher robustness.

\subsection{Audio Quality} \label{sec:audio_quality}

We evaluate audio quality using SNR and ViSQOL, with results shown in Table~\ref{tab:audio_quality}.
From the results, we observe that watermark creation shows a significant difference. As shown in our previous experiments in Figure \ref{fig:watermark_creation}, although AudioMarkBench successfully alters the watermark binary message, the attack lacks robustness.
The watermark creation attack aims to modify the clean binary message into a targeted adversarial message. When the attack prioritizes audio quality, the features of the perturbed audio remain very similar to those of the clean audio, which also implies that the embedded watermark features are relatively weak. Applying a no-box perturbation is more likely to remove these weakened watermark features. Thus, balancing audio quality and attack robustness becomes a critical design consideration. Our previous experiments on the creation attack show that: while our approach yields slightly lower audio quality, it achieves substantially higher robustness.


In the watermark replacement and removal, the audio quality of \ours attack is comparable to that of AudioSeal/Timbre and AudioMarkBench. \ours exhibits the lowest audio quality; however, after the optimization step, its audio quality improves substantially and becomes nearly equivalent to that of AudioSeal/Timbre (without attack).

\begin{table}[t]
    \centering
    \caption{Audio quality comparisons in AudioSeal and Timbre.}
    \label{tab:audio_quality}
    \vspace{-2mm}
    \scalebox{0.56}{
    \begin{tabular}{lllcccc}
        \toprule
        \multirow{2}{*}{Metric} & \multirow{2}{*}{\shortstack{Watermark \\ Method}} & \multirow{2}{*}{Attack Type} & \multicolumn{4}{c}{Method} \\
        \cmidrule(lr){4-7}
        & & & w/o attack & AudioMarkBench & Ours & Ours(+opt) \\
        \midrule
        \multirow{6}{*}{SNR}
        & \multirow{3}{*}{AudioSeal}
            & Replacement & 26.63 & 25.40 & 23.89 & 26.27 \\
        & & Creation    & 26.63 & 43.72 & 25.94 & 37.33 \\
        & & Removal     & 26.63 & 26.49 & 25.49 & 26.59 \\
        \cline{2-7}
        & \multirow{3}{*}{Timbre}
            & Replacement & 25.04 & 25.05 & 24.82 & 25.03 \\
        & & Creation    & 25.04 & 63.85 & 38.19 & 47.84 \\
        & & Removal     & 25.04 & 24.90 & 24.55 & 25.03 \\
        \hline
        \multirow{6}{*}{ViSQOL}
        & \multirow{3}{*}{AudioSeal}
            & Replacement & 4.92 & 4.59 & 4.34 & 4.70 \\
        & & Creation    & 4.92 & 4.83 & 4.26 & 4.65 \\
        & & Removal     & 4.92 & 4.68 & 4.43 & 4.69 \\
        \cline{2-7}
        & \multirow{3}{*}{Timbre}
            & Replacement & 4.66 & 4.65 & 4.39 & 4.56 \\
        & & Creation    & 4.66 & 5.00 & 4.67 & 4.83 \\
        & & Removal     & 4.66 & 4.52 & 4.41 & 4.61 \\
        \bottomrule
    \end{tabular}}
\end{table}

\section{Conclusion}\label{sec:conclusion}


In this work, we analyze watermark decoder outputs and observe that they exhibit approximately normal distributions on benign audio. We show that defenders can leverage this statistical regularity to distinguish benign from attacked signals. To demonstrate the fragility of such defenses, we introduce \ours, an adaptive audio watermark attack. \ours employs a two-stage optimization procedure to evade distribution-based detection while preserving high audio quality. Experiments show that while distribution-based defenses can detect prior attacks, \ours consistently bypasses them, achieving higher attack success rates and lower detection success rates. These findings underscore the urgent need for robust defense mechanisms designed for adaptive attacks on audio watermarking.

\section{Acknowledgments}

This work was supported in part by the U.S. National Science Foundation under Grant CNS-2310207, CNS-2520900, CNS-2451168.

\section{Generative AI Use Disclosure}

The authors used generative AI tools such as ChatGPT solely for grammar improvement and language polishing. All technical contents were conceived, conducted, and verified by the authors, which include the methodology, experiments, analysis, etc. The authors take full responsibility for the content of the publication.

\bibliographystyle{IEEEtran}
\bibliography{bibliography}

\end{document}